\documentclass[11pt]{article}

\usepackage{amsmath}
\usepackage{graphicx}
\usepackage{indentfirst}
\usepackage{amssymb}
\usepackage{cite}
\usepackage{color}
\usepackage{subfigure}

\begin{document}

\title{ The shadows and observational appearance of a noncommutative  black hole surrounded by various profiles of accretions  }

\date{}
\maketitle

\begin{center}
\author{Xiao-Xiong Zeng,}$^{a,b,}$\footnote{E-mail: xxzengphysics@163.com}
\author{Guo-Ping Li,}$^{c,}$\footnote{E-mail: gpliphys@yeah.net}
\author{Ke-Jian He}$^{*,d,}$\footnote{*Corresponding author:kjhe94@163.com}
\vskip 0.25in
$^{a}$\it{State Key Laboratory of Mountain Bridge and Tunnel Engineering, Chongqing\\ Jiaotong University, Chongqing 400074, China}\\
$^{b}$\it{ Department of Mechanics, Chongqing Jiaotong University, Chongqing ~400074, China}\\
$^{c}$\it{Physics and Space College, China West Normal University, Nanchong 637000, China}\\
$^{d}$\it{College Of Physics, Chongqing University, Chongqing 401331, China}\\

\end{center}
\vskip 0.6in
{\abstract
{
The accretion around the black hole plays a pivotal role in the theoretical analysis of black hole shadow, and of the black hole observation in particular. We mainly study the shadow and observation characteristics of noncommutative Schwarzschild black holes wrapped by three accretion models, and  then explore the influence of noncommutative parameters on the observation appearance and spacetime geometry of black holes.  When the black hole is surrounded by an optically and geometrically thin accretion disk, it shows that  the direct emissions always dominate the total observed intensity, while  the lensing ring superimposed upon the direct emission produces a thin ring, which improves the observation intensity of the black hole image. However, the photon rings  makes  a negligible contributions to the total observed brightness due to its exponential narrowness, although the photon ring intersects the thin plane more than three times to pick up larger intensity. More importantly,  when the noncommutative parameters  changed, the corresponding regions and observation intensities of photon ring, lensing ring and direct emission all change correspondingly. For optical thin spherically symmetric accretion, we consider the static and infalling matters, respectively. We find that the observation intensity of the two spherical accretion models increase with the increase of noncommutative parameters. In addition, due to the Doppler effect of the infalling movement, the shadow image of infalling accretion is darker than that of static accretion. In a word, the different accretion models and  noncommutative parameters  will lead to different shadow images and optical appearances of noncommutative Schwarzschild black holes. }
}

\thispagestyle{empty}
\newpage
\setcounter{page}{1}

\section{Introduction}
\label{sec:intro}
The black hole, as one of the important predictions of general relativity, has been widely concerned and  discussed intensively by physicists. In recent years, the observation of gravitational waves  provides a new platform for investigating the properties of black holes \cite{Abbott:2016blz}. In 2015, in the first and second observation runs, the Laser Interferometer Gravitational wave Observatory (LIGO) detector has  identified ten gravitational wave events, which are caused by the merger of binary black holes \cite{LIGOScientific:2018mvr}. In addition, as a major event to observe astrophysical black holes, the Event Horizon Telescope (EHT) collaboration released the first ultra-high angular resolution image of the supermassive black hole M$87^\ast$ in 2019, which  opened  a new window for the  black hole observation\cite{Akiyama:2019cqa,Akiyama:2019brx,Akiyama:2019sww,Akiyama:2019bqs,Akiyama:2019fyp,Akiyama:2019eap,Akiyama:2021qum,Akiyama:2021tfw}. A striking feature of this image is that a dark interior is surrounded by a bright ring, in which the dark area and the bright ring are dubbed  shadow and photon ring of the black hole, respectively. The strong gravitational field near the black hole makes the light deflect, which is widely considered to be an important determinant of the formation of black hole shadow\cite{Bardeen:1972fi,Bardeen:1973tla,Bozza:2009yw}. For the distant observer,  the image of black hole shadow is projected onto the local sky of the observer due to the gravitationally lensed, which is realized by the extremely strong gravitational field around the black hole. Hence, the study of  shadow and gravitational lensing of black hole can be used as a feasible method to explore the  gravitational field\cite{Gibbons:2008rj,Werner:2012rc,Ovgun:2018fnk,Ovgun:2019wej,Javed:2019qyg,Jusufi:2017lsl,Jusufi:2017mav,Li:2019mqw,Li:2020wvn,Pantig:2021zqe,Virbhadra:2007kw,Virbhadra:1999nm,Keeton:2005jd,
Bozza:2002zj,Ovgun:2020yuv,Cunha:2018acu,Javed:2020mjb}.

The study of the shadow of black hole will excavate valuable geometric information near the black hole,  and the observation of black hole shadow can have a deeper understanding of the related properties of black hole. The theoretical analysis of the shadow formation of Schwarzschild black hole began with the work in\cite{Synge:1966okc},  and then the shadow of Kerr black hole was also discussed in\cite {Bardeen:1972fi}.
Additionally in  the universe, the real astrophysical black holes are always surrounded by  various accretion materials, which will play an important role for the observation and imaging of black holes. In 1979, Luminet simulated the first image of a black hole with an emitting accretion thin disk, which shows that the strong gravity of the black hole makes the light around the black hole bend \cite{Luminet:1979nyg}. Although the actual accretion flow is generally not spherically symmetric, the simplified spherical model can give rise to some thoughtful understandings on fundamental properties of accretion appeared in the usual general-relativistic magnetohydrodynamics models\cite{Akiyama:2019fyp,Porth:2019wxk}.  In this sense, by considering the spherically symmetric accretion, the shadow of Schwarzschild black hole has been investigated \cite{Narayan:2019imo}. It is shown that the size of
the observed shadow and photon sphere has nothing to do with the distribution of accretion, which means that shadow is a feature of spacetime geometry and is  unrelated to the details of accretion process.
When considering  the case that the Schwarzschild black hole is surrounded by  an optically and geometrically thin disk accretion, Wald et al. found that the detail and emission form of accretion will affect the appearance and observation intensity of black hole shadow\cite{Gralla:2019xty}. Meanwhile, they found that the black hole image  was composed of the dark region and bright region with inclusion of direct emission, lensing
rings and photon rings, which provide different degrees of help for the observation intensity. The study of black hole shadow has ushered in a new climax, the shadow and observed appearance of the black hole have also been widely studied in general relativity, as well as modified gravity theory, e.g., the size and shape of shadow, the spherical accretion flows, and the thin or thick accretion disks \cite{Falcke:1999pj, Shaikh:2018lcc,Banerjee:2019nnj,Vagnozzi:2019apd,Vagnozzi:2020quf,Safarzadeh:2019imq,Davoudiasl:2019nlo,Roy:2019esk,Chen:2019fsq,Cunha:2019hzj, Konoplya:2020bxa, Roy:2020dyy, Islam:2020xmy, Jin:2020emq, Guo:2020zmf, Wei:2020ght,Caiyifu,Cuadros-Melgar:2020kqn,Li:2021riw,He:2021htq,Konoplya:2019sns,Zhang:2019glo,
Ma:2020dhv,Saurabh:2020zqg,Zeng:2020vsj,Guerrero:2021ues,Gan:2021xdl,Peng:2021osd,Qin:2020xzu,Zhang:2020xub,Peng:2020wun,Zeng:2020dco,Hou:2021okc,Zhang:2021hit,Wang:2021ara,Long:2020wqj,Bambi:2019tjh,Khodadi:2020jij,Johannsen:2010ru, Grenzebach:2014fha,Ovgun:2018tua,Ovgun:2020gjz,Okyay:2021nnh,Ovgun:2021ttv,Gyulchev:2020cvo,Dokuchaev:2019jqq}.

In recent years, noncommutative spacetime in gravity theories has been an important research object \cite{Nicolini:2008aj}, and it is considered to be an alternative way to the quantum gravity\cite{Snyder:1946qz}. In particular, considering the influence of noncommutative on black hole  is a far-reaching subject, and several methods are proposed to implement a noncommutative spacetime in theories of gravity \cite{Li:2020gzi,Aschieri:2005yw,Aschieri:2005zs,Meljanac:2007xb,Meljanac:2006ui,Harikumar:2012zi}. And then, it shows that noncommutativity can be implemented in General Relativity by modifying  the source of matter, so that the Dirac delta function is replaced by a Gaussian distribution \cite{Nicolini:2005vd} or alternatively by a Lorentzian distribution \cite{Nozari:2008rc}.  Several works have been proposed to investigate in the  noncommutative spacetime, e.g., the thermodynamic property, the quasinormal modes, and the characteristics of geometry\cite{Anacleto:2019tdj,Batic:2019zya,Anacleto:2020zfh,Campos:2021sff,Yan:2020hga,Ma:2017jko,Kumar:2017hgs,Ovgun:2019jdo}. However, the observed intensity and appearance characteristics of black holes wrapped by accreted matter have rarely been investigated in the  noncommutative spacetime. Therefore, our aim in this paper is to explore the shadow and optical appearance of noncommutative Schwarzschild black hole, the influence of noncommutative parameters on the shadows and photon rings of black holes. Here, we mainly consider the noncommutative Schwarzschild black holes surrounded by  different accretion models, namely the geometrically and optically thin disk
accretion and the spherically symmetric accretion.  More than that, in the case of thin disk accretion, we will explore the black hole shadow, observation appearance, and corresponding luminosity intensity under different emission functions. In addition, we also explore whether the change of relevant state parameters in noncommutative background will affect the shadows, lensing rings and photon rings of black holes.

This paper is arranged as follows:  Section 2 is devoted to introduce the noncommutative Schwarzschild black hole via Lorentzian distribution and discuss the effective potential and photon orbits of it. In section 3, when an optically thin disk accretion surrounded the black hole,  we give the shadow image of the black hole and its observation intensity. In section 4, we show the images of the black hole with spherically symmetric accretions;
Finally, section 5 ends up with conclusions and discussions.

\section{The effective potential and photon orbit of the noncommutative Schwarzschild black hole }
\label{light}
\label{light}
In this section,  we aim to deduce the photon motion near the noncommutative Schwarzschild black hole, so as to obtain the characteristics of the ray trajectories near the black hole. We consider that the  mass density of a static, spherically symmetric, particle-like gravitational source is no longer a function distribution, but given by a Lorentzian distribution  as follows \cite{Nozari:2008rc},
\begin{equation}
 \rho _{\theta }=\frac{\sqrt{\theta } M}{\pi ^{3/2} \left(\pi \theta +r^2\right)^2}.\label{EQ2.1}
\end{equation}
Here, $ \theta $ is the strength of noncommutativity of spacetime and $M$ is the total mass diffused throughout a region with linear size $\sqrt{\theta }$. For the smeared matter distribution, we can further obtain  \cite{Anacleto:2019tdj},
\begin{equation}
\mathcal{M}_{\theta }=\int_0^r  \rho _{\theta } (r) 4 \pi  r^2  dr=\frac{2 M}{\pi }\left(\tan^{-1}(\frac{r}{\sqrt{\pi  \theta }})-\frac{\sqrt{\pi  \theta } r}{\pi  \theta +r^2}\right)=-\frac{4 \sqrt{\theta } M}{\sqrt{\pi } r}+M+ \mathcal{O}(\theta ^{3/2}). \label{EQ2.2}
\end{equation}
In this way, the noncommutative Schwarzschild black hole metric is given by
\begin{equation}
d s^2=-A(r) d t^2+\frac{1}{A (r)}d r^2+r^2 \left(d \vartheta ^2+\sin ^2  \vartheta d \varphi^2   \right),\label{EQ2.3}
\end{equation}
and
\begin{equation}
A (r)=1-\frac{2 \mathcal{M}_{\theta }}{r}=1-\frac{2 M}{r}+\frac{8 \sqrt{\theta } M}{\sqrt{\pi } r^2}+\mathcal{O}(\theta ^{3/2}).\label{EQ2.4}
\end{equation}
The event horizon of this line element can be got at $A (r)=0$, and the larger root corresponds to the event horizon of the black hole and the smaller root is so called Cauchy (or inner) horizon, which is
\begin{equation}
r_e=M+\frac{\sqrt{M^2 \pi-8 \sqrt{\pi } \sqrt{\theta } M}}{\sqrt{\pi }},\label{EQ2.5}
\end{equation}
and
\begin{equation}
r_c=M-\frac{\sqrt{ M^2 \pi-8 \sqrt{\pi } \sqrt{\theta } M}}{\sqrt{\pi }}.\label{EQ2.6}
\end{equation}
In the next step, we study  the motion of photons near the black hole. In this case, the motion of photon is described by the Euler-Lagrange  equation in the background (\ref{EQ2.3}), given by
\begin{equation}
\frac{d}{d\tau }\left(\frac{\partial \mathcal{L}}{\partial \dot{x}^{\mu }}\right)=\frac{\partial \mathcal{L}}{\partial x^{\mu }}.\label{EQ2.7}
\end{equation}
Here, $\tau$ is the affine parameter and  $\dot{x}^{\mu }$  represents the four-velocity of light ray. In the context of static spacetime, we can get the Lagrangian  $\mathcal{L}$  of photons, that is,
\begin{equation}
\mathcal{L}=\frac{1}{2} \mathit{g}_{\mu \nu } \dot{x}^{\mu } \dot{x}^{\nu }=\frac{1}{2} \left(-A (r) \dot{t}^2+A (r)^{-1} \dot{r}^2+r^2 \left(d \vartheta ^2+\sin ^2  \vartheta d \varphi^2   \right)\right).\label{EQ2.8}
\end{equation}
The spherical symmetry allows us to choose a particular plane in  spacetime, which for convenience is chosen to be the equatorial plane with  $\vartheta =\frac{\pi }{2}$
\cite{Gralla:2019drh,Synge:1966okc,Bardeen:1972fi}. Obviously,  the Lagrangian  is independent of both  coordinates $t $ and $\varphi$, and we have two conserved quantities
\begin{align}
E=-\frac{\partial \mathcal{L}}{\partial \dot{t}}=A (r) \dot{t},\quad
J=\frac{\partial \mathcal{L}}{\partial \dot{\varphi} }=r^2 \dot{\varphi}.\label{EQ2.9}
\end{align}
Here, $E$ and $ J$ are the energy and angular momentum of photons, respectively. The constraint on $\mathit{g}_{\mu \nu } \dot{x}^{\mu } \dot{x}^{\nu }=0$ (null geodesics) for the line element therefore reads as
\begin{align}
-A (r) \dot{t}^2+\frac{\dot{r}^2}{A (r)}+r^2 \dot{\varphi} ^2=0.\label{EQ2.10}
\end{align}
Further, using the equations (\ref{EQ2.9}) and (\ref{EQ2.10}) we get
\begin{equation}
\dot{r}^2=E^2-\frac{ J^2 }{r^2}A(r). \label{EQ2.11}
\end{equation}
At this point, the effective potential $V_{eff} (r)$ is used to represent the radial geodesic write as follows
\begin{equation}
\dot{r}^2+V_{eff} (r)=\frac{1}{b},\label{EQ2.12}
\end{equation}
and
\begin{equation}
V_{eff} (r)=\frac{1}{r^2}(1-\frac{2 M}{r}+\frac{8 \sqrt{\theta } M}{\sqrt{\pi } r^2}).\label{EQ2.13}
\end{equation}
In which, $b$ is the impact parameter $b=L/E$, it is defined as the vertical distance between the geodesic line and the parallel line passing through the origin. In addition, the affine parameter $\tau$ has been replaced by $\frac{\tau }{\left| J\right| }$. In this spacetime, there exists one null geodesic on the equatorial plane, which is circular. This is essentially the photon sphere, projected on the equatorial plane, yielding a circle, known as photon circular orbit. For the spherically symmetric black hole with a photon sphere, the position of the maximum of the potential corresponds to the stability threshold for the circular null geodesic around the structure $b_c$. When $b > b_c$, the existence of $dr/d\varphi=0$, that is, there is a turning point for the null geodesic, and if there is no inflection point, the light will be captured by the black hole. For the critical case $b = b_c$, the photon is in the critical state of being captured or escaping from the black hole. At this time, the photon will move to an unstable circular orbit near the black hole, and the corresponding surface of the circular orbit is the surface of the photon sphere. At the same time, the motions of the light ray on the  sphere plane should have a definite value
\begin{equation}
\dot{r}=0, \quad  \ddot{r}=0.\label{EQ2.14}
\end{equation}
Therefore, we have
\begin{equation}
V_{eff} (r)=\frac{1}{b^2}, ~~~ V_{eff} ^{'}(r) = 0.\label{EQ2.15}
\end{equation}
Based on this equation, the effective potential of the black holes exhibits a maximum for the photon sphere radius $r_p$ corresponding to the real and the positive. Therefore,  the radius $r_p$ and impact parameter $b_c$ of photon sphere satisfy
\begin{align}
r_p{}^2-b_c^2 A(r)=0,\quad 2 b_c^2 A(r)^2-r_p{}^3 A'(r)=0.\label{EQ2.16}
\end{align}
Because of the change of correlation parameters, the spacetime structure will be different, which means that the motion behavior of photon is different. Therefore, we give a table of various important physical quantities under the different strength of noncommutativity $\theta$ in this spacetime, which is presented in Table 1.
\begin{table}[h]
\caption{The values of the relevant physical quantity for different state parameter $\theta$  which $M = 1$.}
\begin{tabular}{| c | c | c | c | c |c | c | c |   }
\hline
Parameter &  $\theta=0.001$  & $\theta=0.005$  & $\theta=0.01$   & $\theta=0.02$   & $\theta=0.03$  & $\theta=0.04$   & $\theta=0.049$\\
\hline
 $r_e$       &1.92589 &1.82513 &1.74071     &1.60141      &{1.46716}    &{1.31192}   &{1.02984} \\
\hline
$r_{p}$  	&2.90162  &2.76952 & 2.66073    &2.48667      &{2.32854}    &{2.16678}   &{2.00178}  \\
\hline
$b_{c}$     &5.06897  &4.89998 &4.76285    &4.54807       &{4.35966}    &{4.17583}   &{4.00178}  \\
\hline
\end{tabular}
\label{table1}
\end{table}

When the value of noncommutative parameter $\theta$ increases, it is easy to see that the radius of the event horizon  $r_e$ and the radius of the photon sphere $r_p$ decrease, and the impact parameter $b_c$  of photon sphere also decreases. As a cross validation of this result, we can compare the results in Table 1 with those in Schwarzschild spacetime. We know that in Schwarzschild spacetime, the radius of the photon sphere is $r_p \sim 3M$ (the Schwarzschild spacetime ). From Table 1, it goes to show that the radius of the photon sphere is smaller  than that in the  Schwarzschild spacetime, and the difference is more obvious with the increase of  the strength of noncommutativity $\theta$.  Concurrently, taking $\theta=0.01$ and $\theta=0.049$ as examples, we show the effective potential $V_{eff}$ (or impact parameter $b$) versus the radius $r$ in Figure 1.
\begin{figure}[tbp]
\centering 
\includegraphics[width=.45\textwidth]{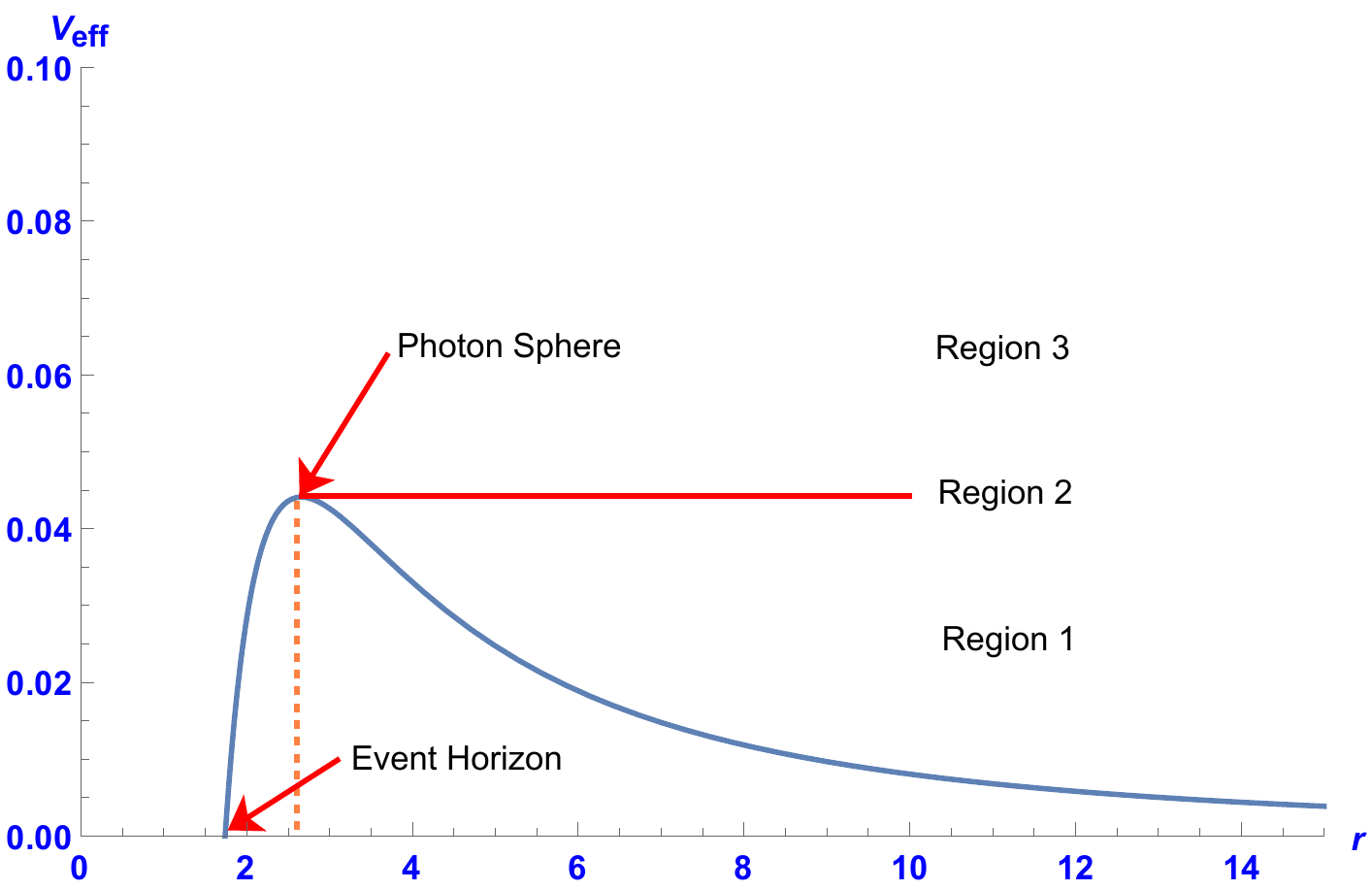}
\hfill
\includegraphics[width=.45\textwidth]{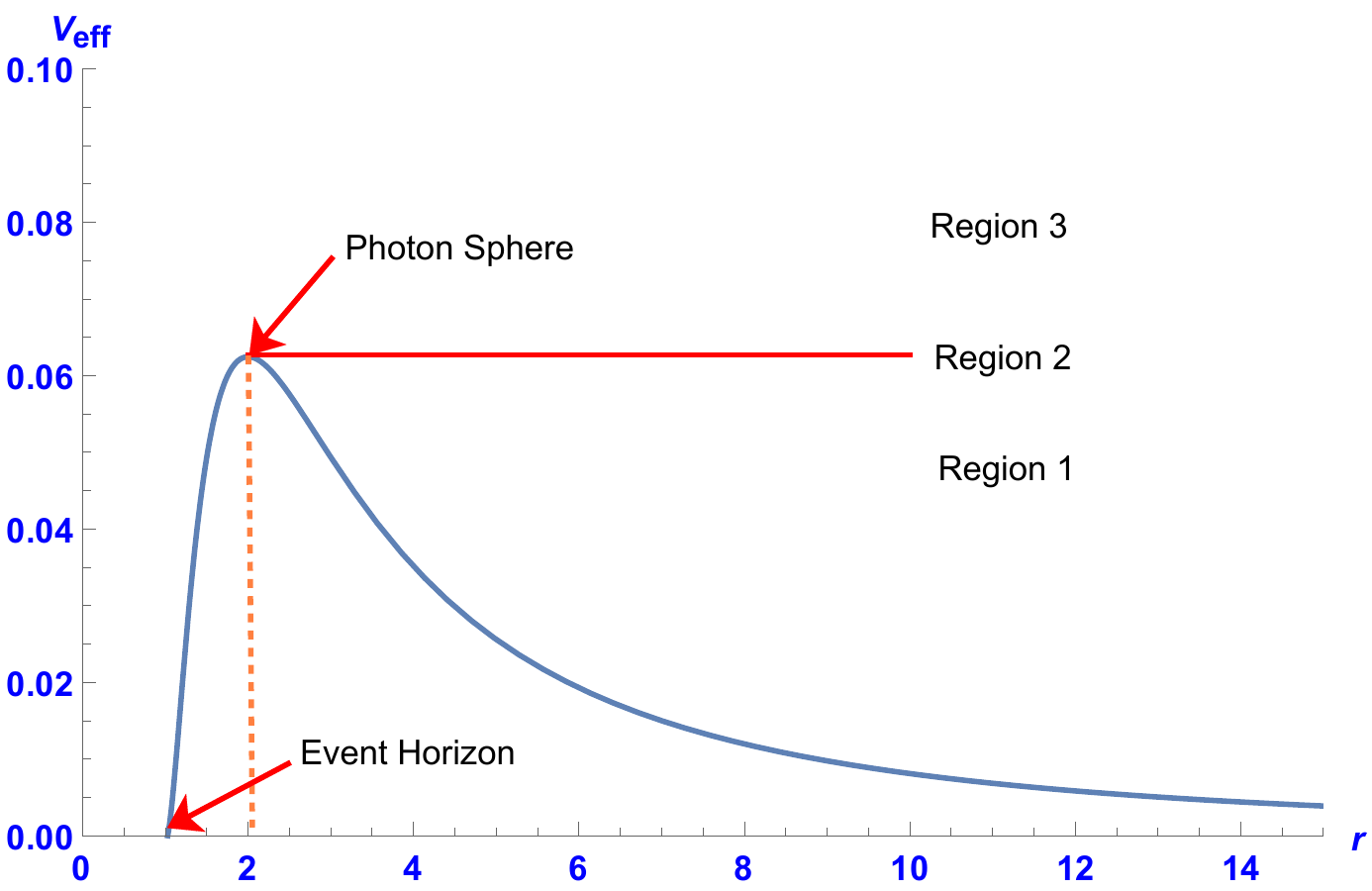}
\caption{\label{fig1}   The profiles of the effective potential $V_{eff}$  and $r$ for $\theta=0.01$ (left panel) and $\theta=0.049$ (right panel) with $M=1$. }
\end{figure}
When $V_{eff}=0$, it corresponds to the position of event horizon. With the increase of the radius $r$, the effective potential increases, and the maximum value is reached in the position of the photon sphere, i.e $r\sim b_c$.
After that, the effective potential begins to decrease with the increase of the radius $r$, i.e $r>b_c$. In general, when a light ray  moves in the radially inward direction, the light ray will show different motion behavior. The reflection occurs when light rays hit the potential barrier generated by the effective potential, which corresponds to Region 1 ($b>b_c$). And, when the impact parameter of the light ray is close to the radius of the photon sphere, the light ray  will rotate around the black hole, which corresponds to  Region 2 ($b=b_c$). If there are no obstacles to affect the behavior of light rays, it will fall into the black hole, which corresponds to Region 3 ($b<b_c$). After that, we are going to introduce a parameter $u_0=1/r$ so that we can rewrite the equation (\ref{EQ2.12}) as
\begin{align}
 \Phi(u_0)=\frac{du_0}{d\varphi }=\sqrt{\frac{1}{b^2}-u_0^2 \left(1-2 M u_0+\frac{8 \sqrt{\theta } M}{\sqrt{\pi } u_0^{-2} }\right)}.\label{EQ2.11}
\end{align}
By solving the above equation numerically and with the help of the ray-tracing model, we can obtain the behavior of the geodesic lines for different values of the impact parameter $b$. In  Figure 2, taking $\theta = 0.01$ (left panel) and $\theta = 0.049$ (right panel) as two examples, we show the ray trajectories near the black hole.
\begin{figure}[h]
\centering 
\includegraphics[width=.4\textwidth]{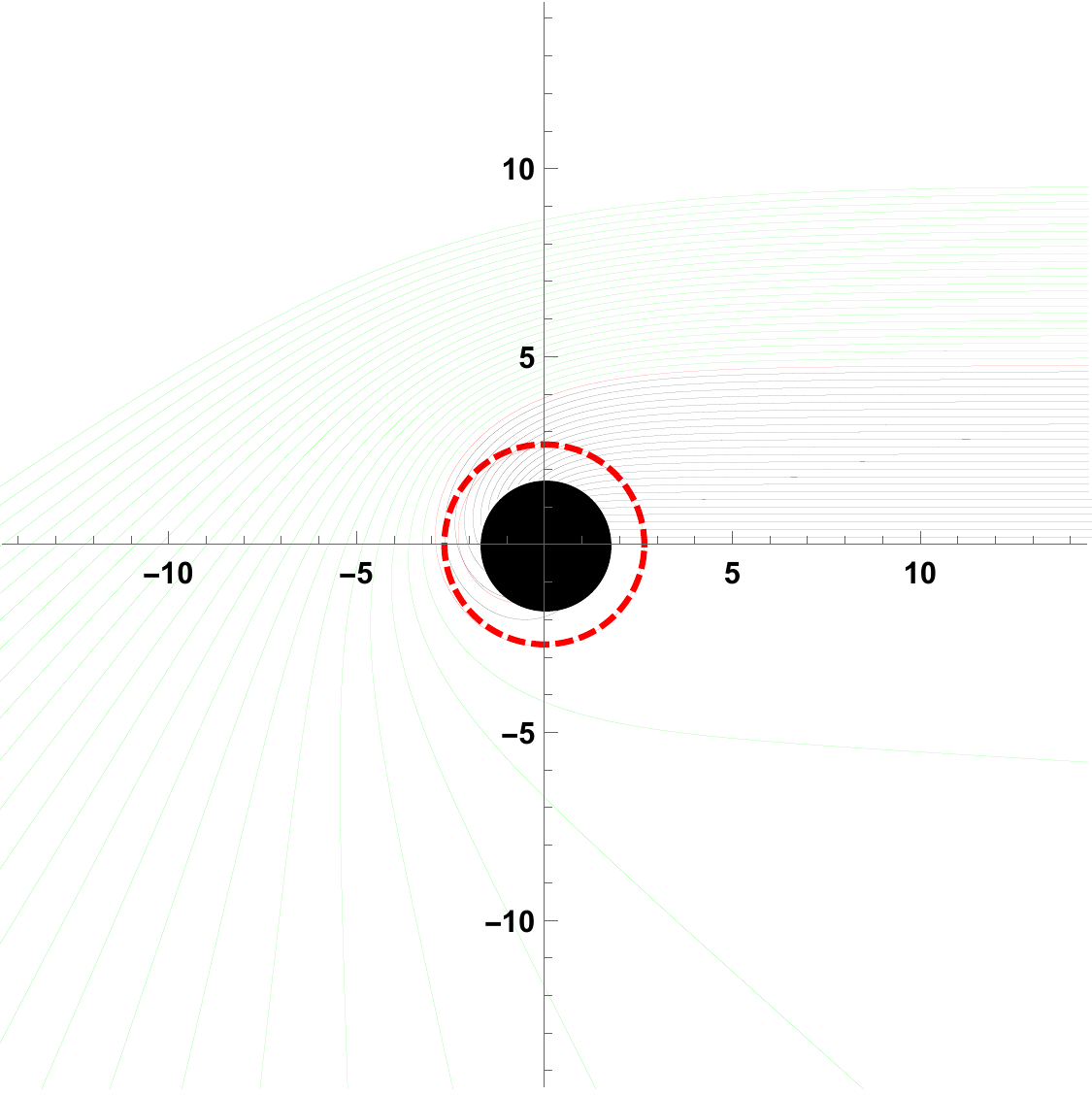}
\hfill
\includegraphics[width=.4\textwidth]{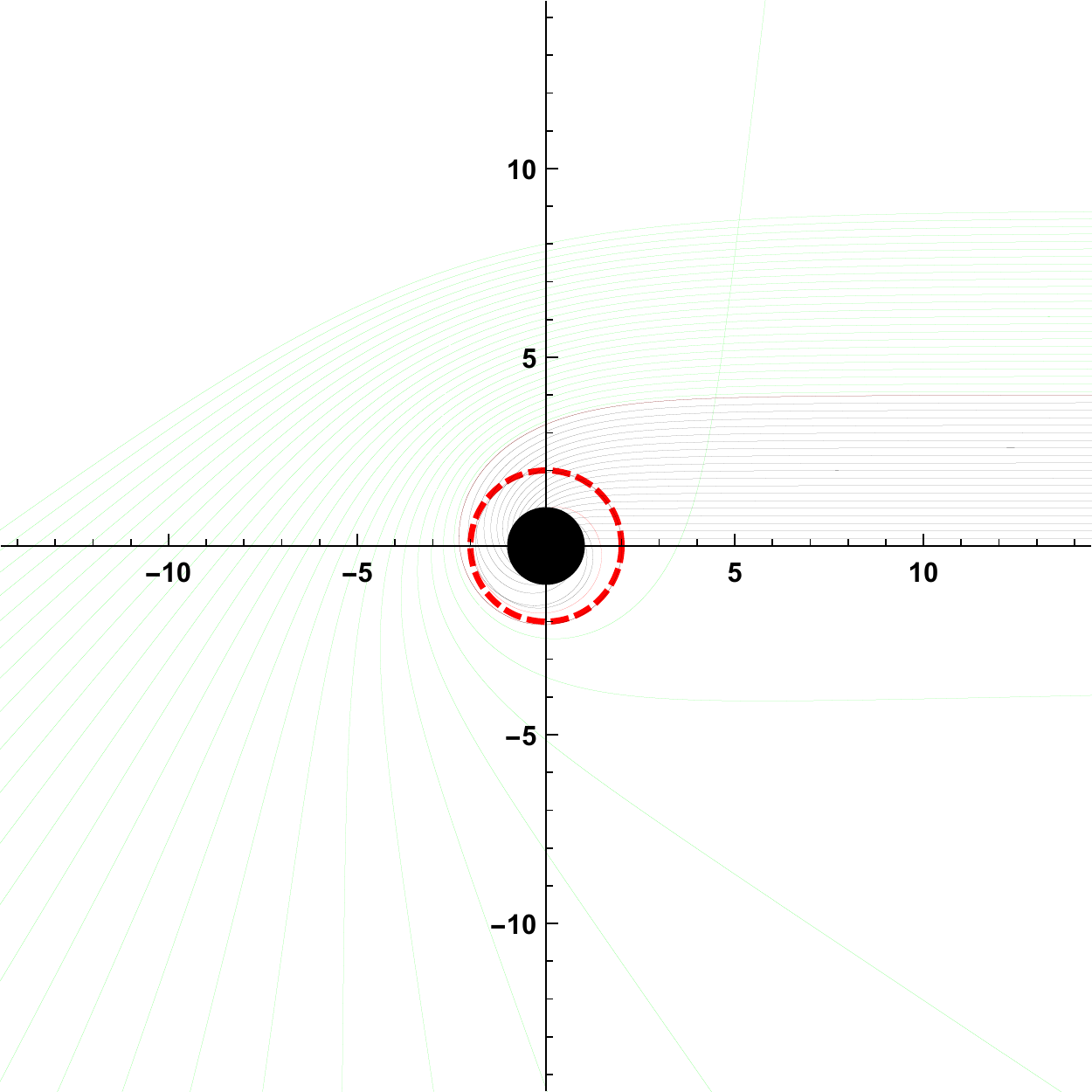}
\caption{\label{fig2}  The trajectory of the light ray  in the polar coordinates $(r, \varphi)$ with different  noncommutative parameter values, where  $\theta=0.01$ (left panel) and $\theta=0.049$ (right panel). }
\end{figure}

In Figure 2, we have a black disk representing the limit of the event horizon, and the red dotted circle represents the location of the  photon sphere. We show the change of trajectory of light ray under different impact parameter $b$, and here we set the distance between different impact parameter $b$ to $1/5$. When $b < b_c$ (the black line in Figure 2), light will eventually be sucked into the black hole, and can not be captured by the observer at infinity. When $b > b_c$ (green line in Figure 2), the light deflects so that the light can reach the observer at an infinite distance. Especially in the critical case of $b= b_c$ (the red line in Figure 2), the light will revolve around the black hole. Moreover, the green line, red line and black line in Figure 2 correspond to region 1, region 2 and region 3 in Figure 1, respectively. In addition, due to the change of noncommutative parameter, the size of black hole shadow and  the binding ability to photons will change accordingly, which also leads to the different deflection degree of each ray path in Figure 2.


\section{Shadows, photon rings and lensing rings with thin disk accretions }
It is well known that there are always many accretion materials around black holes in the real universe. Nonetheless, considering many factors involved in the analysis of black hole shadow in the noncommutative spacetime, such as potential geometric background, geometry, optics and emission of accretion disk model, it is useful to consider some simplified assumptions in order to study the observational characteristics of black hole shadow.
In this section, we consider an optically and geometrically thin disk around the equatorial plane of the black hole in noncommutative spacetime, and the observed light intensity is mainly provided by the thin disk.
\subsection{Direct emission, Lensed ring and photon ring}
The light ray trajectories near the black hole are an important basis for studying the observational appearance of a black hole. Meanwhile, in order to obtain the optical appearance of the noncommutative black hole through the ray-tracing procedure, we define the total number of orbits generated by a beam of light from the light source to the observer as  the change of azimuthal angle, that is,
\begin{equation}
n (b)=\frac{\varphi }{2 \pi },\label{EQ3.1}
\end{equation}
which is a function of impact parameter $b$. The number of orbits obviously depends on how close the impact parameters $b$ are to the critical parameters $b_c$. In addition, it also depends on the geometry of noncommutative spacetime. In the previous study of Wald et al.\cite{Gralla:2019xty}, they found that the number of photon orbits can be divided into three regions, namely
\begin{itemize}
\item $n<3/4$ : Direct emission, which means that the trajectories of light rays will intersect with the equatorial plane only once.
\item $3/4<n<5/4$: Lensing ring, which means that the trajectories of light rays will intersect with the equatorial plane at least twice.
\item $n>5/4$: Photon ring, which means the trajectories of light rays will intersect with the equatorial plane at least three times.
\end{itemize}
It is worth mentioning that we  use $M = 1$ as the unit, and the numerical regions of  the direct emission, lensing ring and photon ring related to impact parameter $b$ are given under different noncommutative parameter values $\theta$, as shown in  Table 2.
\begin{table}[t]
\caption{Under different noncommutative parameters, the region of direct emission ring, lensing ring and photon ring is related to the impact parameters $b$ for the case  $M=1$.}
\begin{tabular}{| c | c | c | c | }
\hline
Parameter       & $\theta=0.01$                & $\theta=0.03$           & $\theta=0.049$ \\
\hline
Direct emission &$b<4.55053$                   &$b<4.08502$              &$b<3.45242$  \\
 $n<3/4$        &$ b>5.80841$                  &$b>5.50959$              &$b>5.28931$   \\
\hline
Lensing ring  	&$4.55053<b<4.75069$           & $4.08502<b<4.33838$     &$3.45242<b<3.92753$ \\
$3/4 < n < 5/4$ &$4.80392<b<5.80841 $          &$4.41739<b<5.50959$      &$4.09183<b<5.28931$  \\
\hline
Photon ring     &$4.75069<b<4.80392$           &$4.33838<b<4.41739$      &$3.92753<b<4.09183$  \\
$n > 5/4$       &                  &             &                                             \\
\hline
\end{tabular}
\label{table2}
\end{table}
Further, we want to show the direct emission, lensing ring and photon ring with different noncommutative parameters more intuitively. Therefore, we take different values of $\theta$ and show their comparative results in Figure 3. Meanwhile, the corresponding photon trajectories of the  noncommutative spacetime in the polar coordinates ($b$, $\varphi$), are shown in  Figure 4.
\begin{figure}[h]
\centering 
\includegraphics[width=0.65\textwidth]{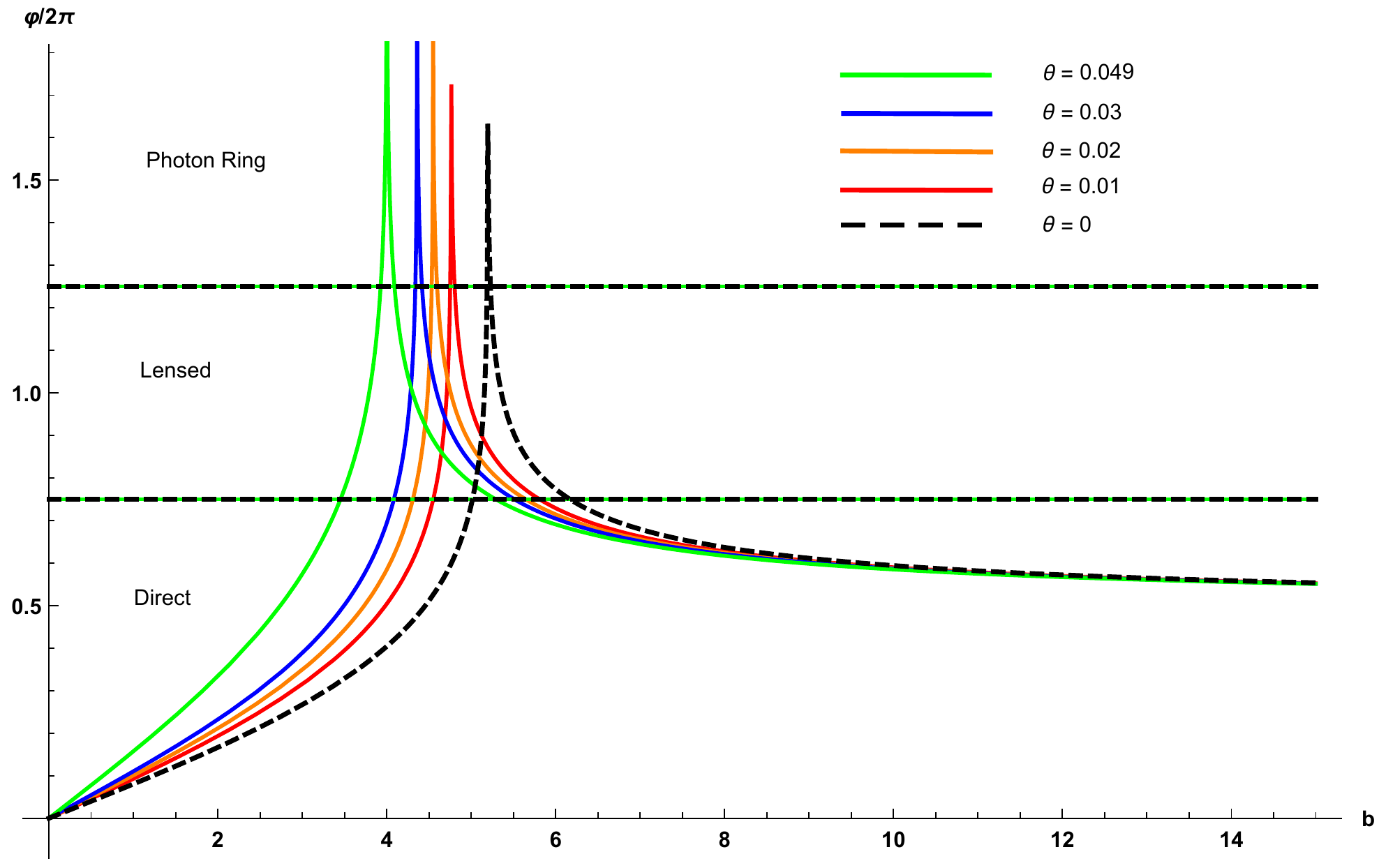}
\caption{\label{fig3}  The number of orbits $n$ versus the impact parameter $b$ for different noncommutative parameter values $\theta$,.  }
\end{figure}

\begin{figure}[h]
\centering 
\includegraphics[width=.325\textwidth]{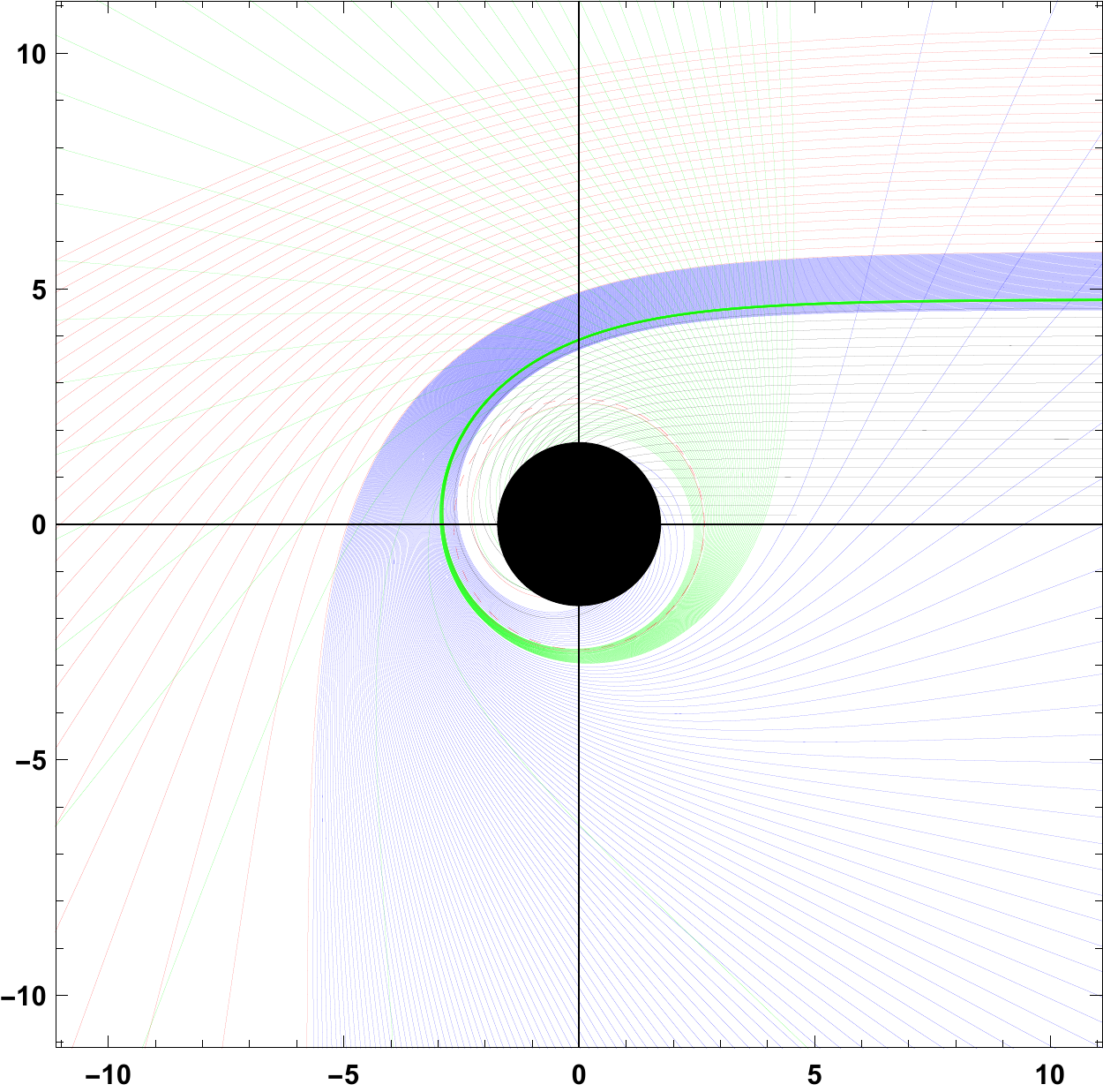}
\hfill
\includegraphics[width=.325\textwidth]{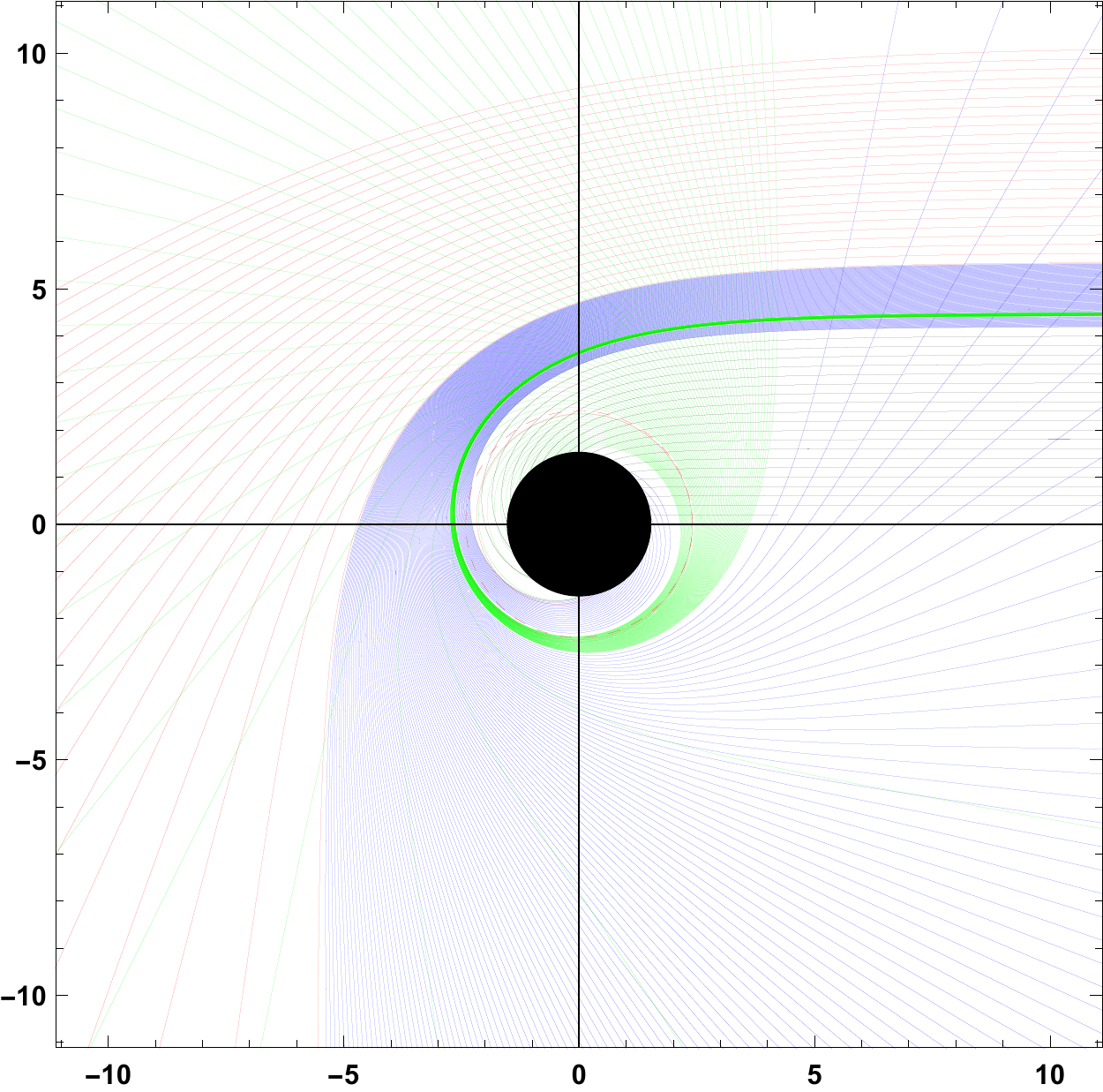}
\hfill
\includegraphics[width=.325\textwidth]{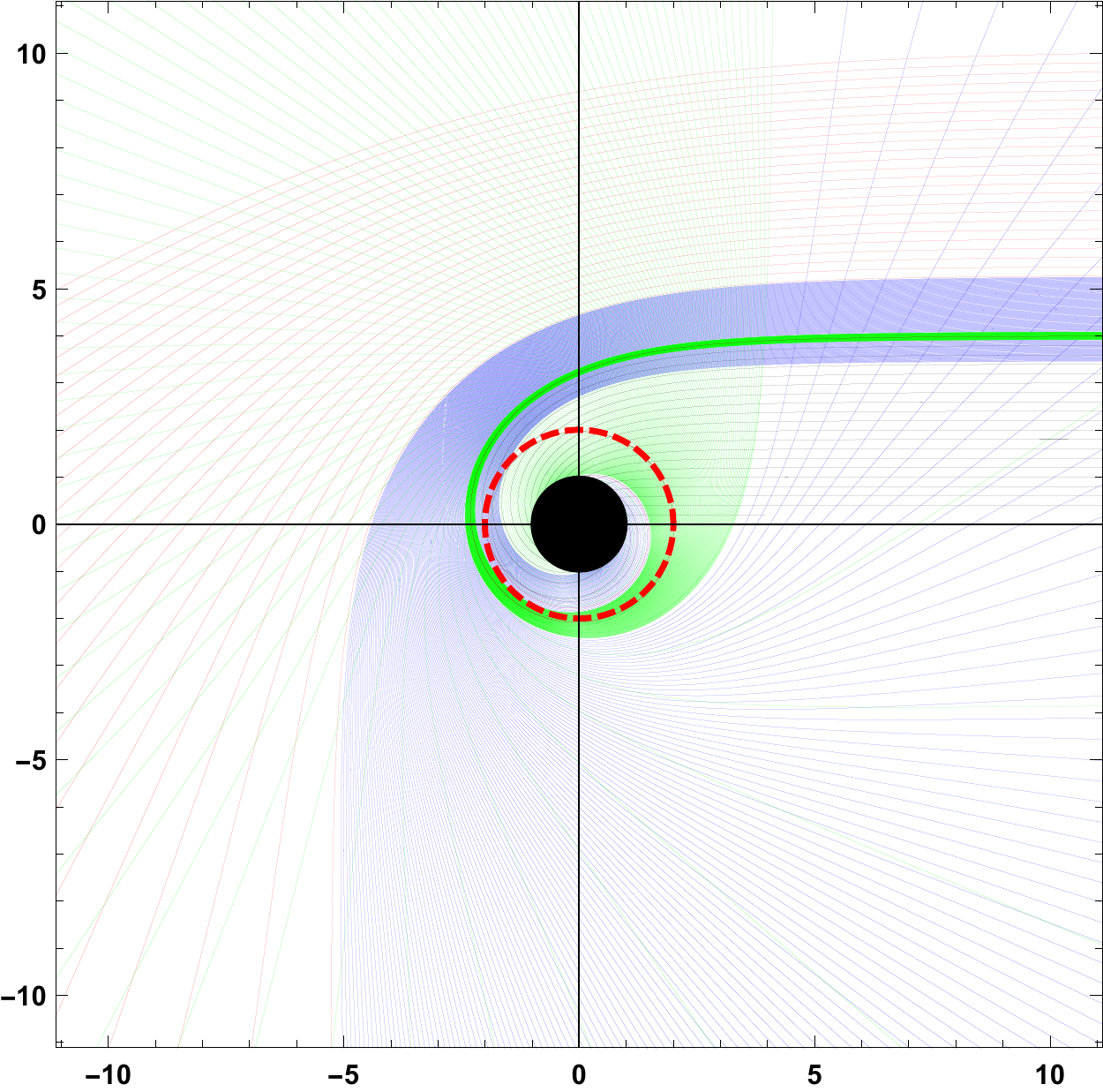}
\caption{\label{fig4} The relationship between the photon trajectories and impact parameter $b$ near the black hole for different values of noncommutative parameters $\theta$. In which, $\theta=0.01$ (left panel), $\theta=0.03$  (mid panel) and $\theta=0.049$ (right panel). The spacings of $b$ are $1/5$, $1/100$, $1/1000$ for the direct (redline), lensed (blue line) and photon (green line) ring trajectories, respectively. And, the red dotted line represents the photon orbit, and the black disk represents the event horizon of the black hole. }
\end{figure}

Combined with Table 2, Figure 3 and Figure 4, we can notice that with the increase of noncommutative parameter $\theta$, the  range  of  lensing ring and photon ring are enlarged.  In other words, with the increase of noncommutative parameters, there will be wider lensing rings and photon rings in spacetime, and the corresponding brightness of these regions will also be enhanced. In addition, when the impact parameter $b$ is close to the critical case $b\sim b_c$, the total number of photon orbits $n$ will reach a peak in a narrow region. After that, if the impact parameter $b$ increases to a large enough value, the light ray trajectories are the  direct emissions no matter what the value of noncommutative parameter $\theta$ is.

\subsection{Transfer functions}
Then, we are in position to study the image of a  black hole with an optically and geometrically thin disk accretion in the noncommutative spacetime, and the disk is located on the equatorial plane of the black hole.
For convenience, the thin disk accretion is located in the rest frame of static worldlines, and the photons emitted from it should follow the basic property of isotropy.
In addition, we consider that the static observer locates at the north pole, and the observed intensity is mainly provided by the thin disk, that is, the influence of other light sources in spacetiime is ignored.

In this way, the specific intensity and frequency of the emission are expressed as $I_{\nu }^{{em}} (r) $ and $\nu_{em}$,  and the observed specific intensity and frequency are defined as $I_{\nu'}^{{obs}} (r)$ and  $\nu_{obs}$.  The observed specific intensity can be expressed as
\begin{equation}
I_{\nu ' }^{{obs}}= [1-\frac{2 M}{r}+\frac{8 \sqrt{\theta } M}{\sqrt{\pi } r^2}]^{3/2} I_{\nu }^{{em}}(r).\label{EQ3.2}
\end{equation}
Then, by integrating the full frequency, we can get the expression of total observation and emission intensity in the whole path, namely
\begin{equation}
I_{{OBS}}=\int  I_{\nu ' }^{{obs}} (r) \, d\nu_{obs} ' =\int  [1-\frac{2 M}{r}+\frac{8 \sqrt{\theta } M}{\sqrt{\pi } r^2}]^2 I_{\nu }^{{em}} (r) \, d\nu_{em} =[1-\frac{2 M}{r}+\frac{8 \sqrt{\theta } M}{\sqrt{\pi } r^2}]^2 I_{{EM}} (r),  \label{EQ3.3}
\end{equation}
where $I_{{EM}}(r)=\int I_{\nu }^{{em}} (r) d \nu_{em}$. In addition, $I_{OBS}$ and $I_{EM}$ represent the total observed and emission intensity, respectively. Based on the previous consideration, the thin disk provides the main contribution of the observation intensity, that is, when the light ray trajectories pass through the thin disk once, it will get brightness from the thin disk once. Finally, the observation intensity that the observer at infinity can obtain should be the total brightness that the light can obtain from the thin disk. Hence, the total received luminosity will be the sum of all the intensities from all these crossings with the accretion disk, which is
\begin{equation}
I_{{OBS}}(r)=\sum _n (A (r))^2 I_{{EM}}(r)|_{r= r_n(b)}. \label{EQ4.4}
\end{equation}
Here, $r_n(b)$ is defined as transfer function, which contains the information about the radial position of the $n_{th}$ intersection with the thin emission disk. In addition,   the slope of the transfer function is $dr/d b$, which is called the  demagnification factor, and it  represents the demagnified scale of the transfer function\cite{Gralla:2019xty,Zeng:2020vsj}.
\begin{figure}[h]
\centering 
\includegraphics[width=.325\textwidth]{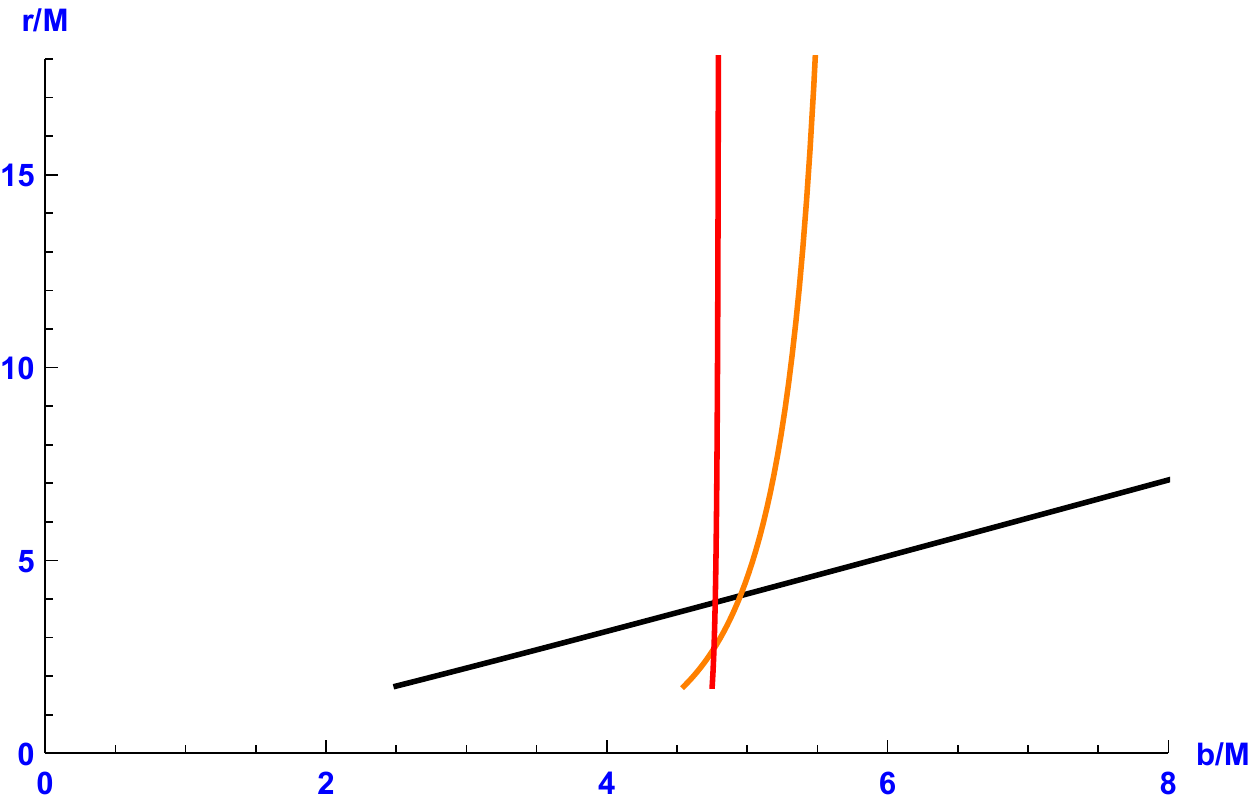}
\hfill
\includegraphics[width=.325\textwidth]{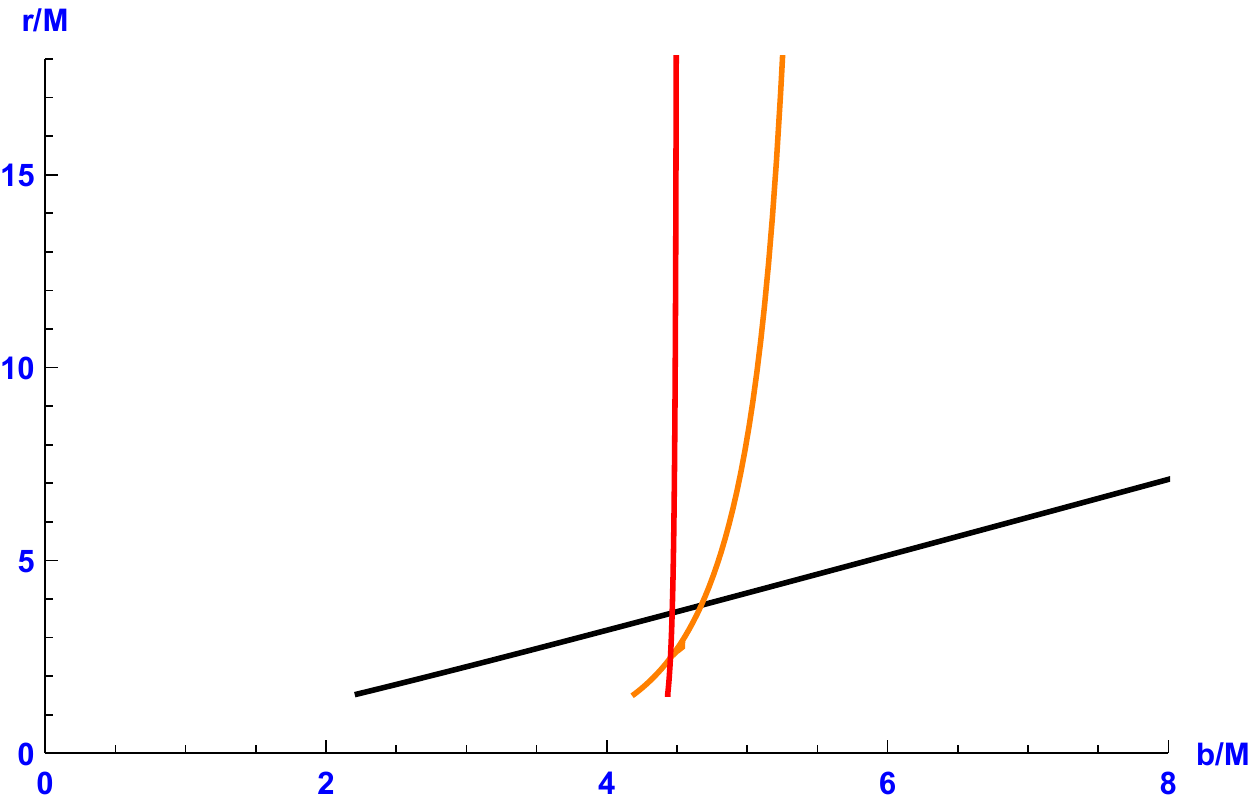}
\hfill
\includegraphics[width=.325\textwidth]{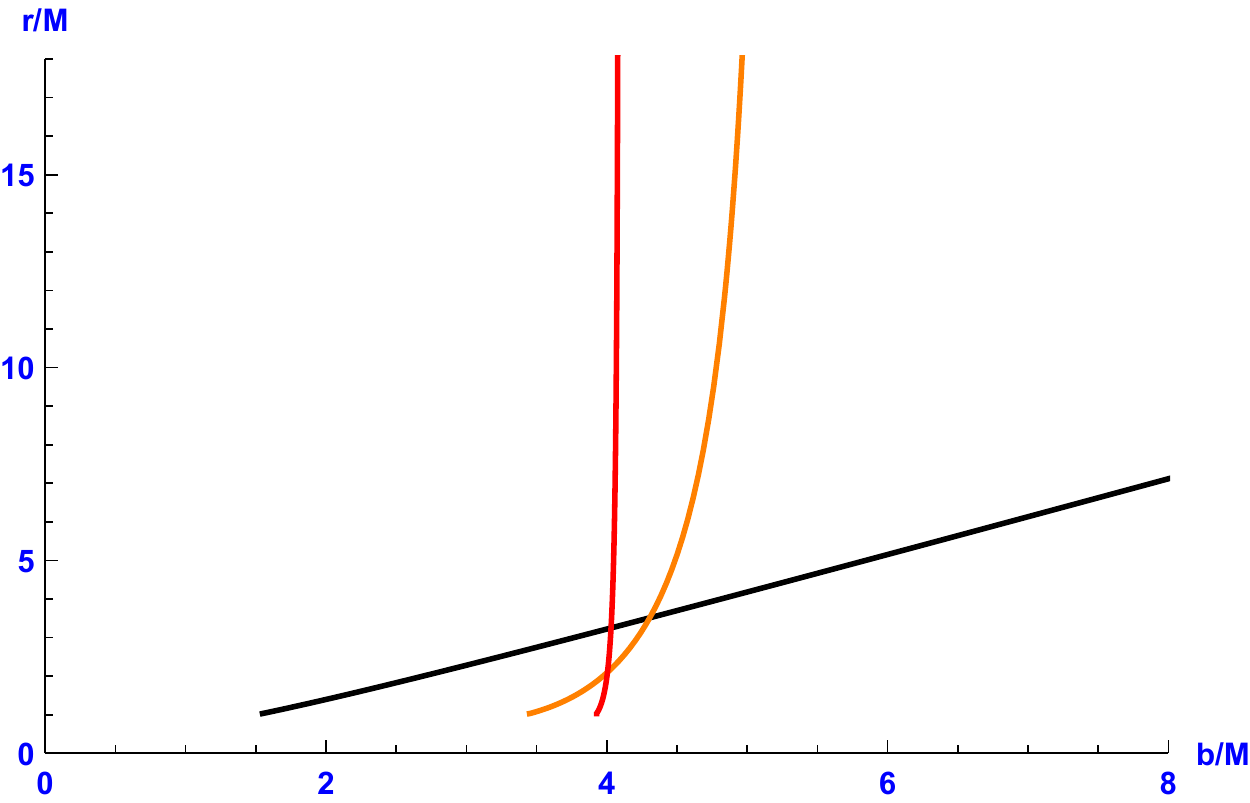}
\caption{\label{fig5}  The first three transfer functions of a black hole under different noncommutative parameters. Here, $\theta = 0.01$ (left panel), $\theta =
0.03$(mid panel) and $\theta = 0.049$ (right panel). }
\end{figure}

In Figure 5, the black line (first transfer function $n=1$ ), which corresponds to the transfer function for the direct emission. Although the value of noncommutative parameter changes, the slope of the black line  is  $\frac{d r}{d b}|_1\sim Constant$. In fact, the direct image profile is the redshifted source profile since its slope is approximatively equal to $1$.
The orange line (second transfer function $n=2$ ),  which corresponds to the transfer function for the  lensing ring. It can be seen that when the impact parameter $b$ is close to the critical case $b\sim b_c$, there is a relatively gentle area, that is, the slope of the transfer function will have a relatively small value. However, as the value of impact parameter $b$ continues to increase, the slope of the second transfer function begins to increase sharply to a very large value. In other words, the image of the back side of the disk will be demagnified, because its slope is much greater than $1$.
The red line ( third transfer function $n=3$), which corresponds to the transfer function for the  photon ring. It is obvious that the  slope $\frac{d r}{d b}|_3$ is close to infinity $\frac{d r}{d b}|_3 \sim \infty$.
Namely, the image of the front side of the disk will be extremely demagnified.
Hence, the total observed flux is mainly provided by the direct emission, and the luminosity provided by photon ring and lensing ring only accounts for a very small part of the total observed flux.
In particular, the image provided by the later transfer function is more demagnetized and negligible, so we only consider the first three transfer functions.

\subsection{The observational features of noncommutative Schwarzschild black hole }
With the previous preparation, we can further study the observed specific intensity.
For the observer, the observed light intensity is mainly provided by the thin disk, and the specific brightness only depends on the radial coordinate $r$. Therefore, we consider  three toy models about the emissivity of thin disks.
\begin{itemize}

\item Model I: we consider that the  emission starts from the position of the radius of the innermost stable circular orbit $r_{isco}$, and the emitted function $I_{EM}(r)$ is a decay function suppressed by the second power, which is
\begin{align}
    I_{EM}^1(r) =\begin{cases}\left(\frac{1}{r-(r_{isco}-1)}\right)^2, &  r>r_{{isco}}  \\
    0, &r \leq r_{{isco}} \label{EQ3.5}
    \end{cases}
\end{align}
\item Model II: we consider that the emission has an obvious peak on the photon sphere layer, which has the similar center and asymptotic characteristics as model I. But the attenuation is greater, so that $ I_{EM}(r)$ is a decay function suppressed by the third power function, which is described by
\begin{align}
    I_{EM}^2(r) =\begin{cases}\left(\frac{1}{r-(r_{p}-1)}\right)^3, &  r> r_{p}   \\
    0, &r \leq r_{p} \label{Eq3.6}
    \end{cases}
\end{align}

\item Model III: we consider that the  emission starts right off the event horizon, but its decay rate is slower than the first two models, as follows
\begin{align}
    I_{EM}^3(r) =\begin{cases} \frac{1-\tan^{-1}(r-(r_{isco-1)}))}{1-\tan^{-1}(r_p)}
    , &  r>r_e  \\
    0,  &r \leq r_e  \label{Eq3.7}
    \end{cases}
\end{align}

\end{itemize}
In this case, by using those emitted functions and with the help of equation (\ref{EQ3.3}), we take $\theta = 0.01$ and $\theta=0.049$ as two examples to show the results of  intensity of the emission and observation in Figure 6 and Figure 7.

In Figure 6 ($\theta = 0.01$) and Figure 7 ($\theta=0.049$),  the left column is the emission profile of the three modes, which is the relationship between the emission specific intensity and the radial coordinate $r$, and from top to bottom are model I, model II and model III, respectively. The middle column shows the observed specific intensity $I_{OBS}$, which is related to the impact parameter $b$. In addition, the optical appearance of observed intensities is given in the column on the right.
\begin{figure}[h]
\centering 
\includegraphics[width=.35\textwidth]{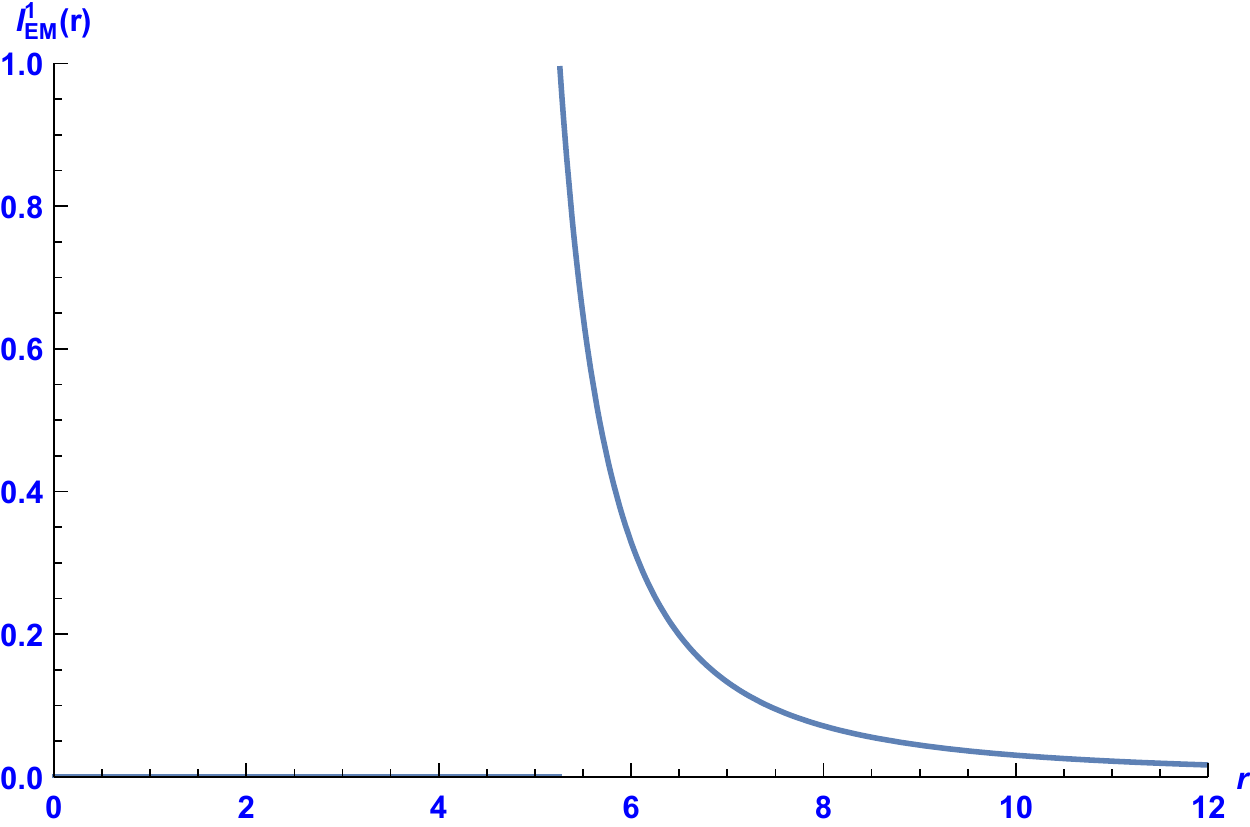}
\includegraphics[width=.35\textwidth]{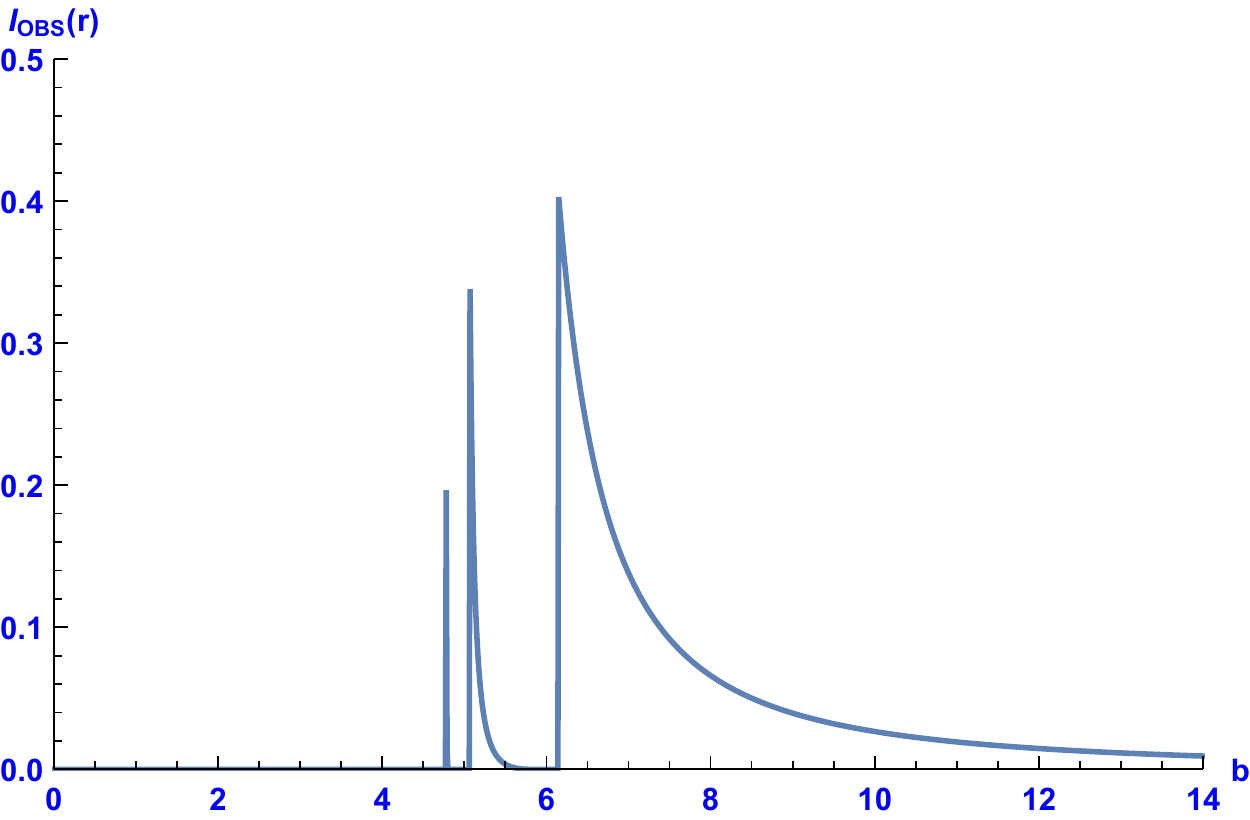}
\includegraphics[width=.28\textwidth]{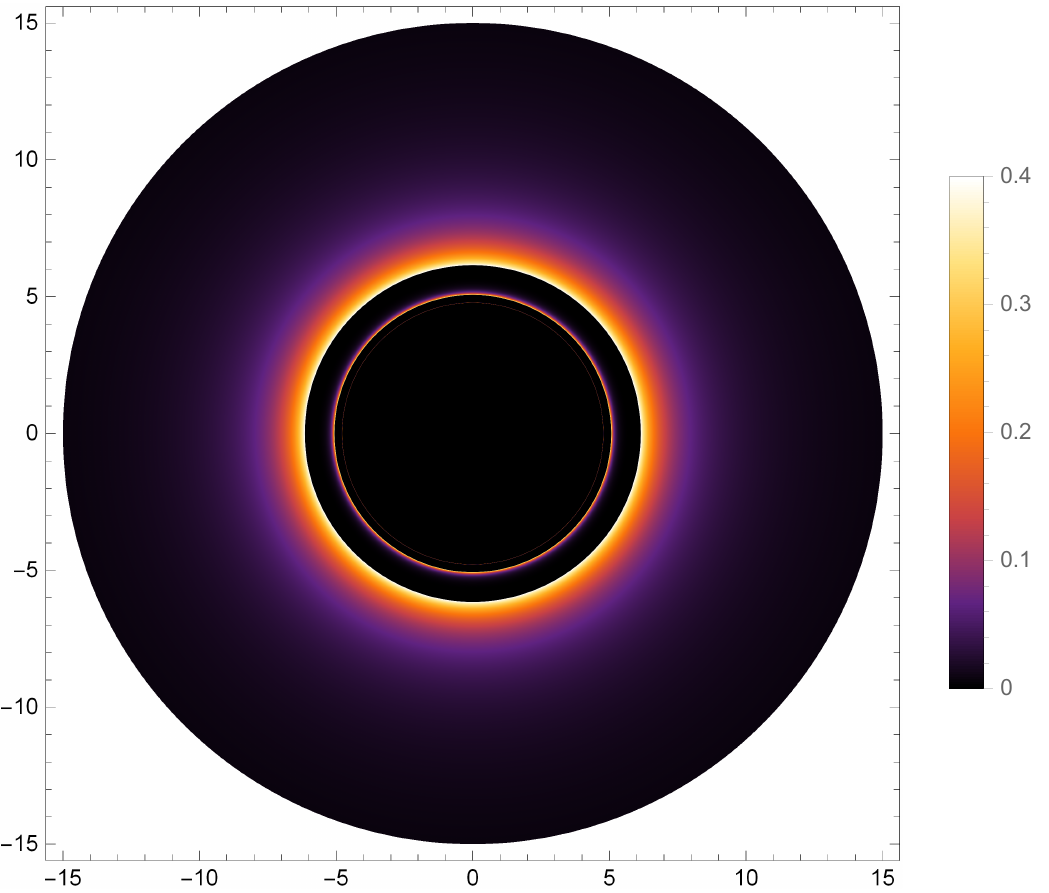}
\includegraphics[width=.35\textwidth]{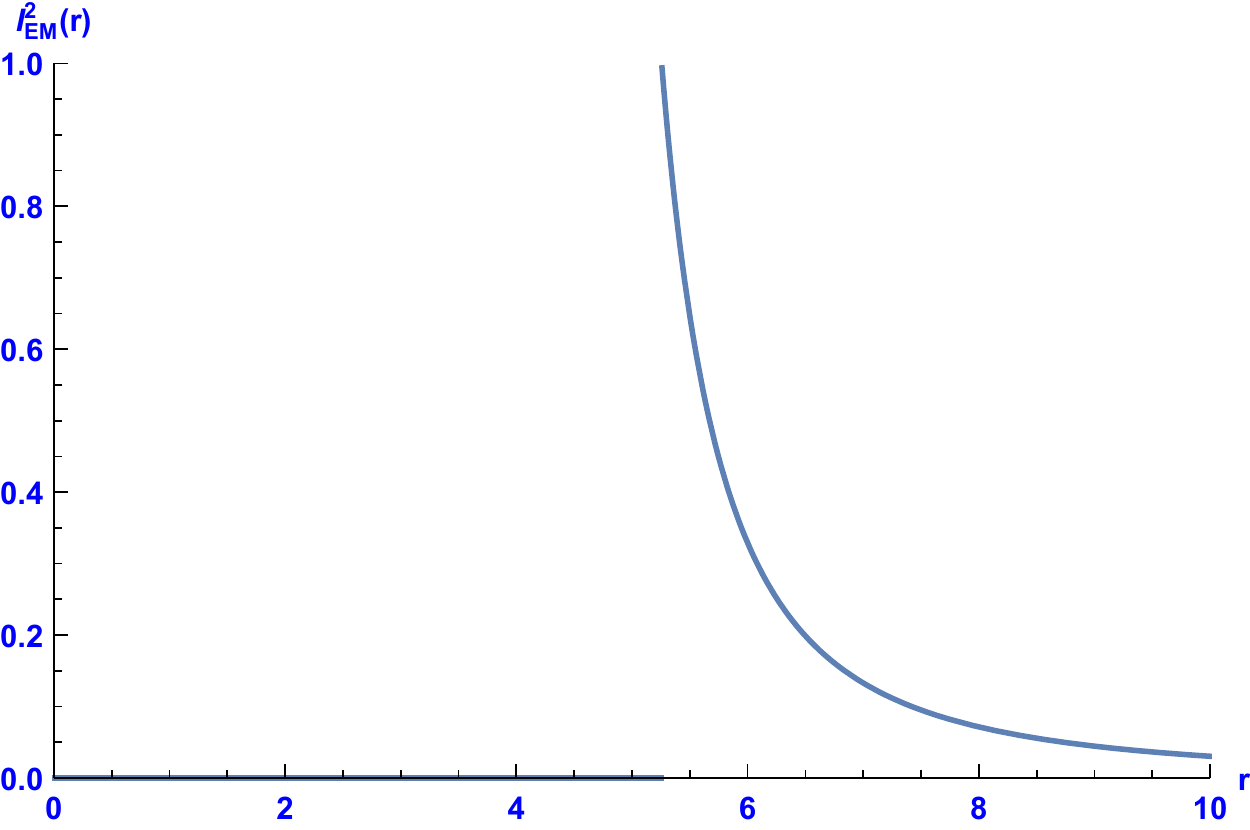}
\includegraphics[width=.35\textwidth]{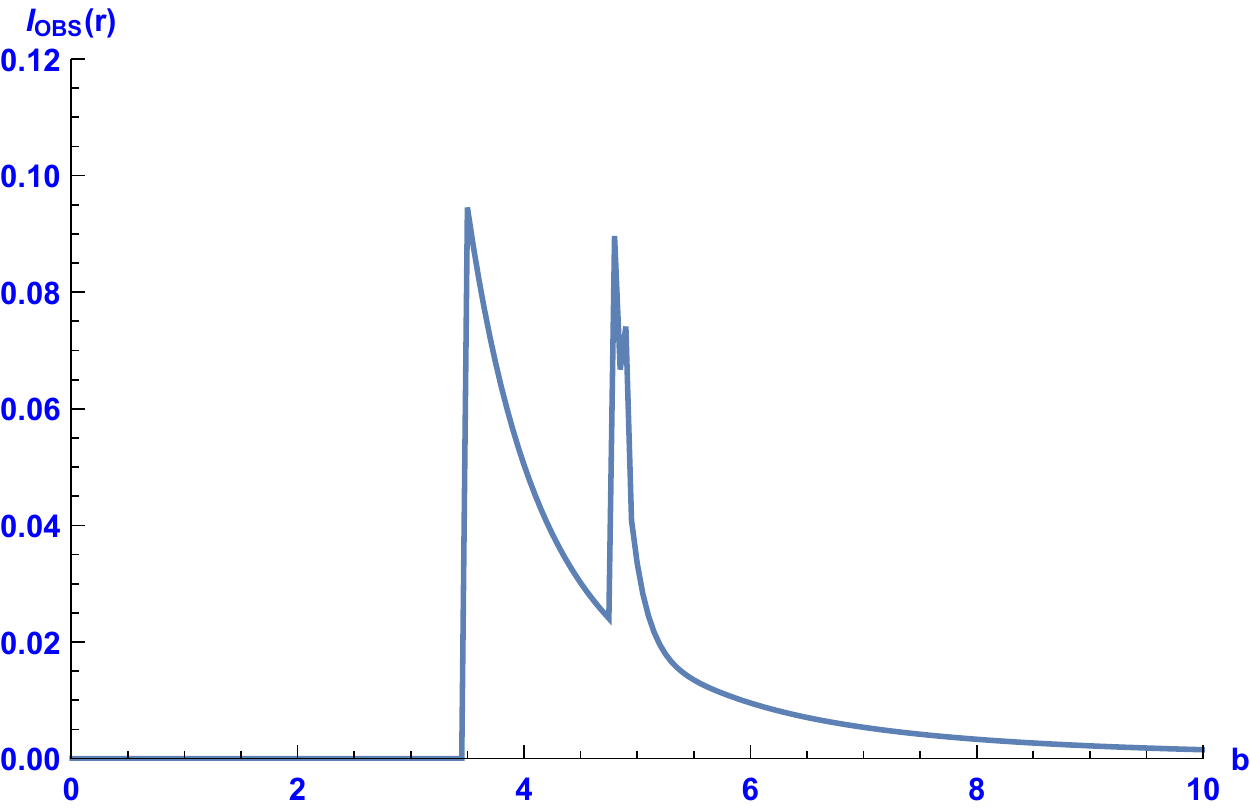}
\includegraphics[width=.28\textwidth]{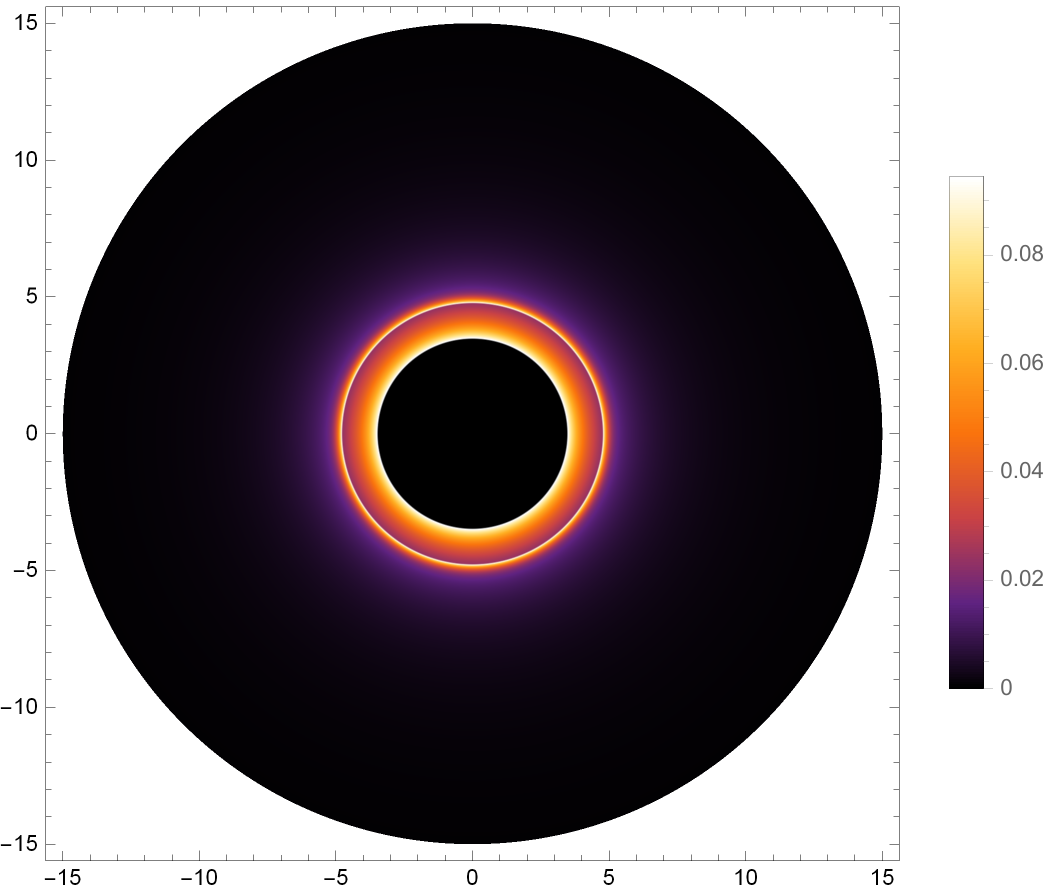}
\includegraphics[width=.35\textwidth]{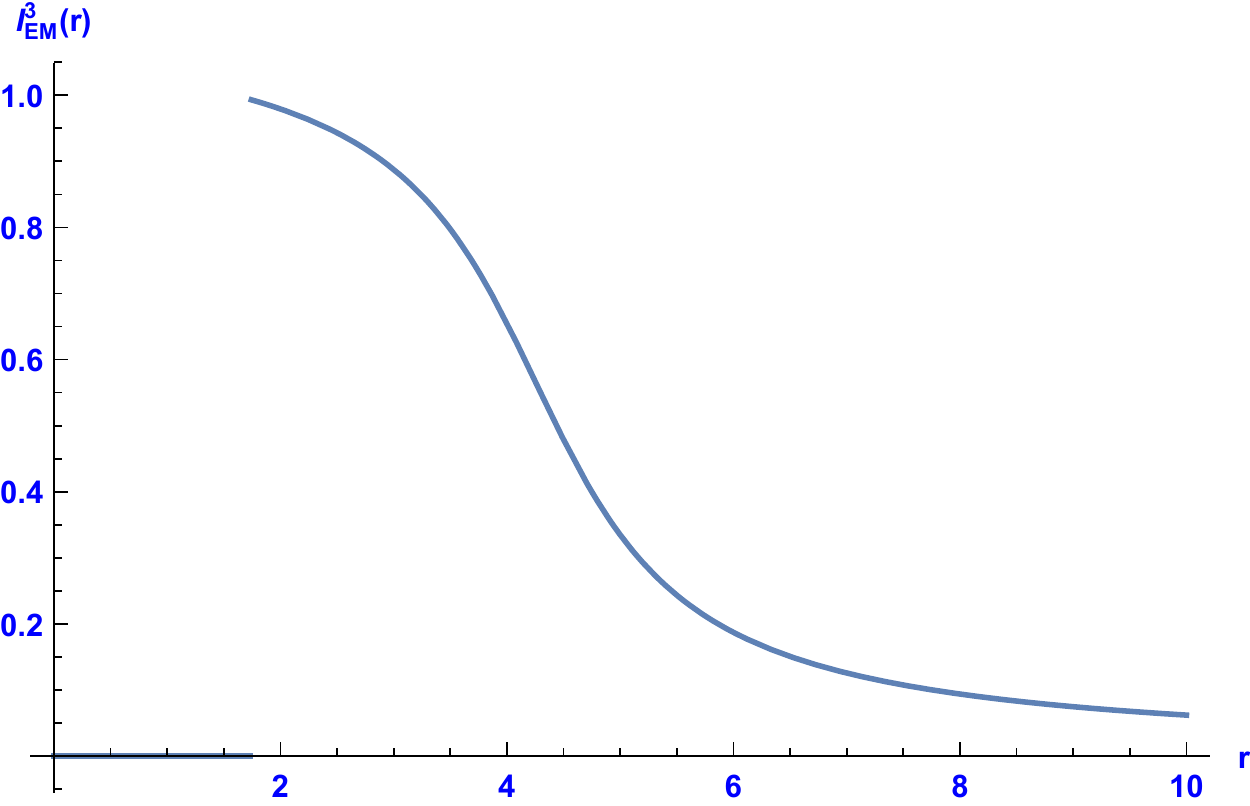}
\includegraphics[width=.35\textwidth]{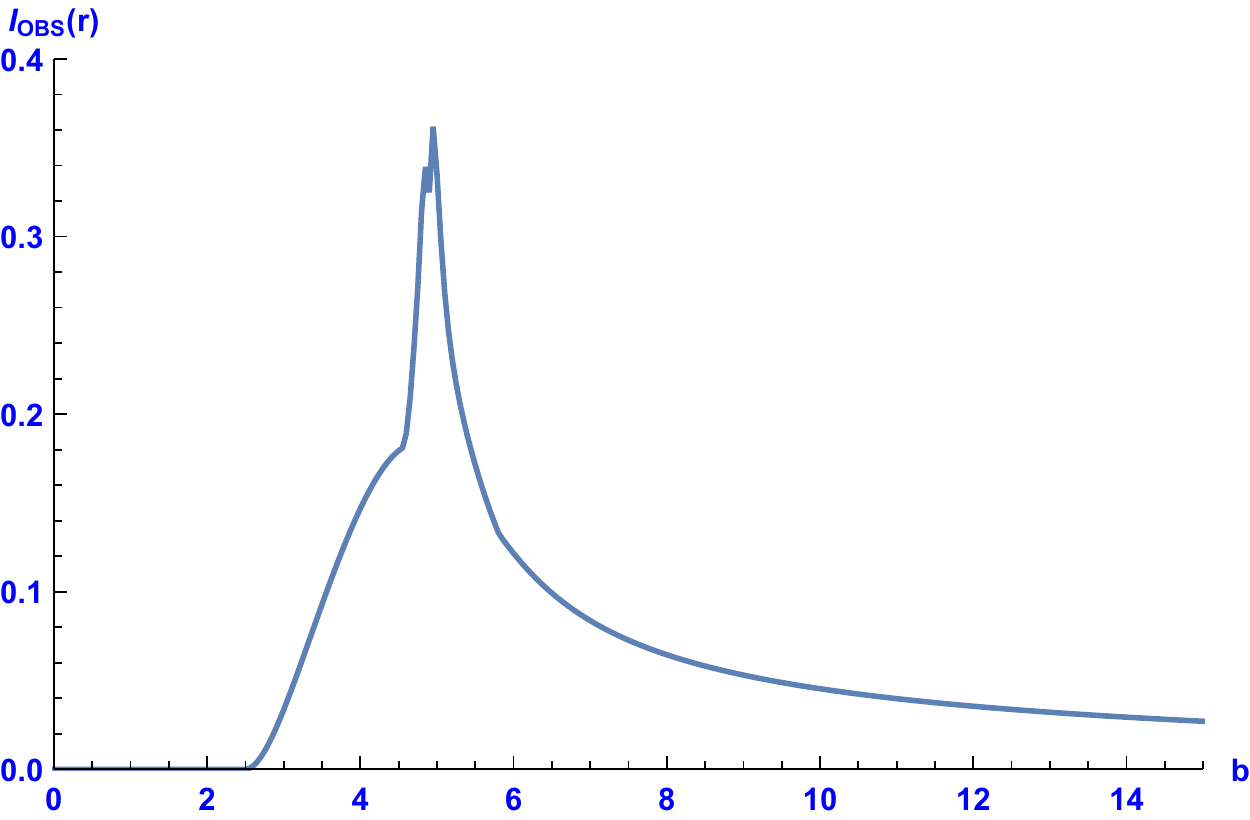}
\includegraphics[width=.28\textwidth]{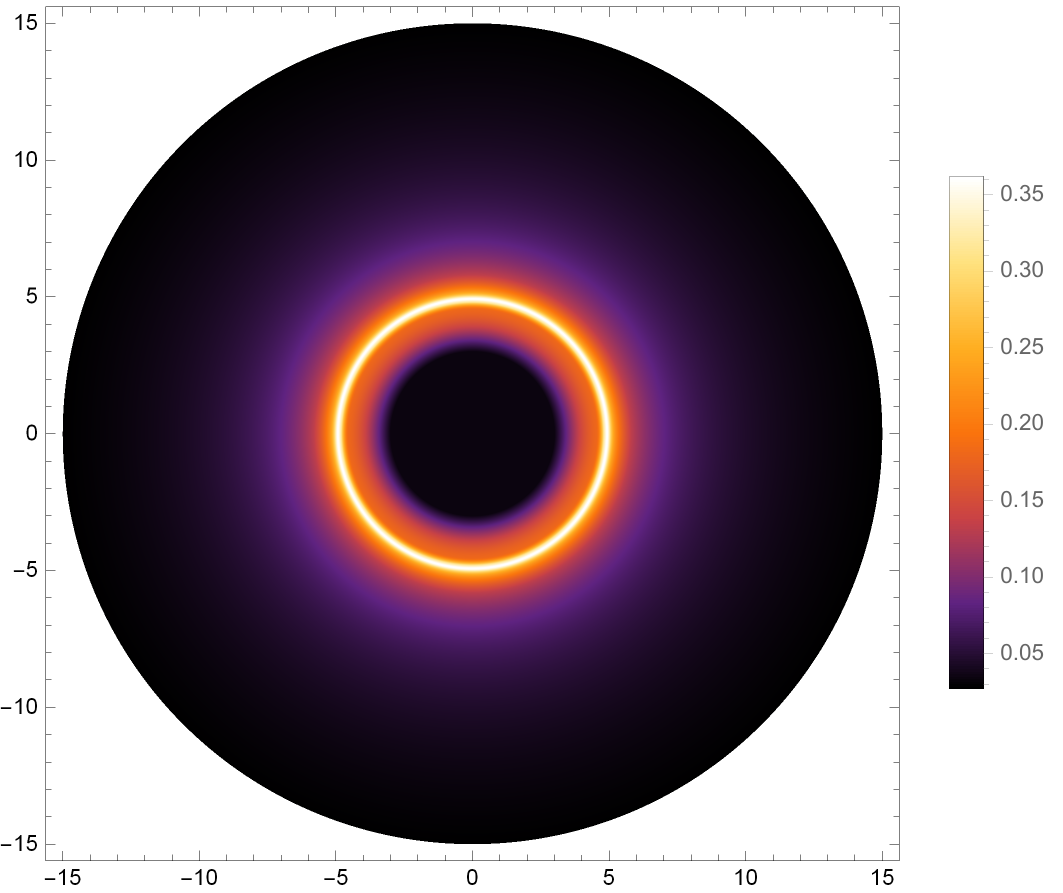}
\caption{\label{fig6}  Observational appearance of the thin disk with different emission profiles  near the black hole ($\theta = 0.01$), viewed from a face-on orientation.}
\end{figure}
\begin{figure}[h]
\centering 
\includegraphics[width=.35\textwidth]{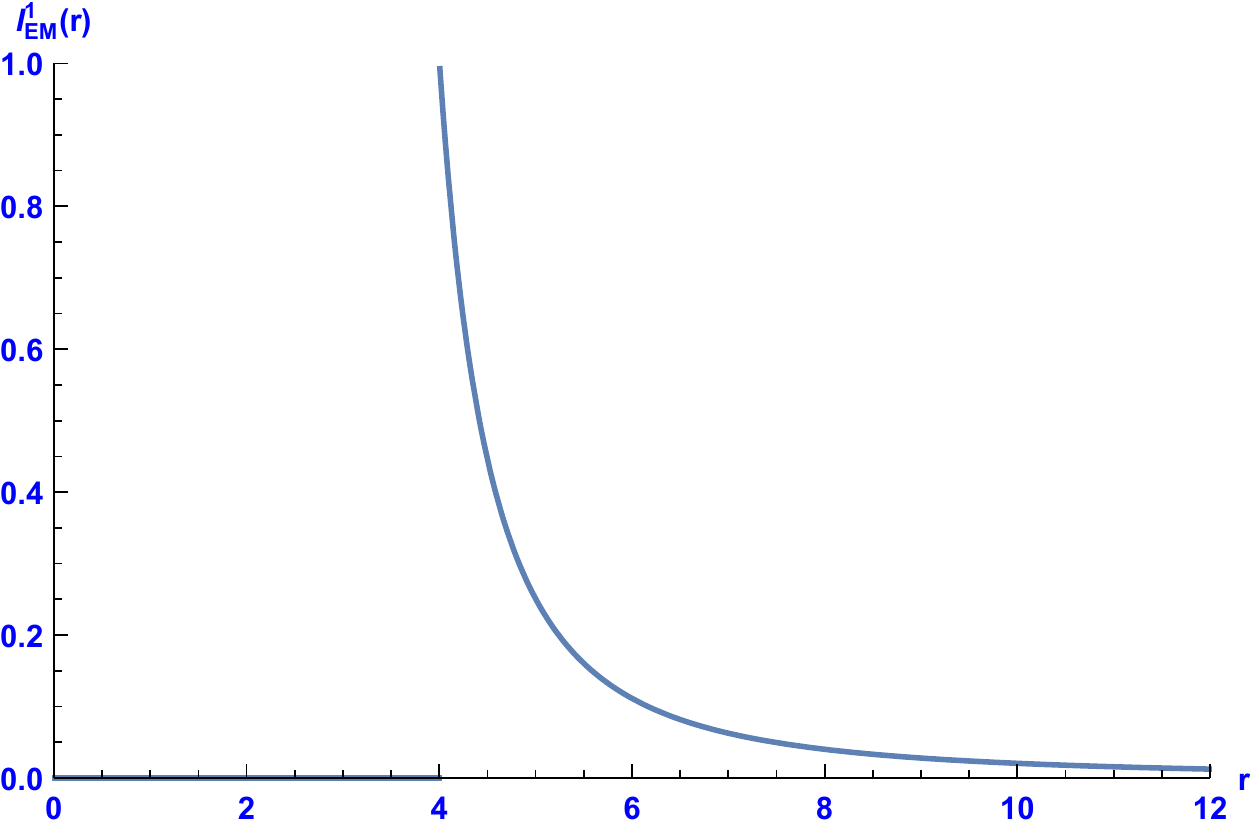}
\includegraphics[width=.35\textwidth]{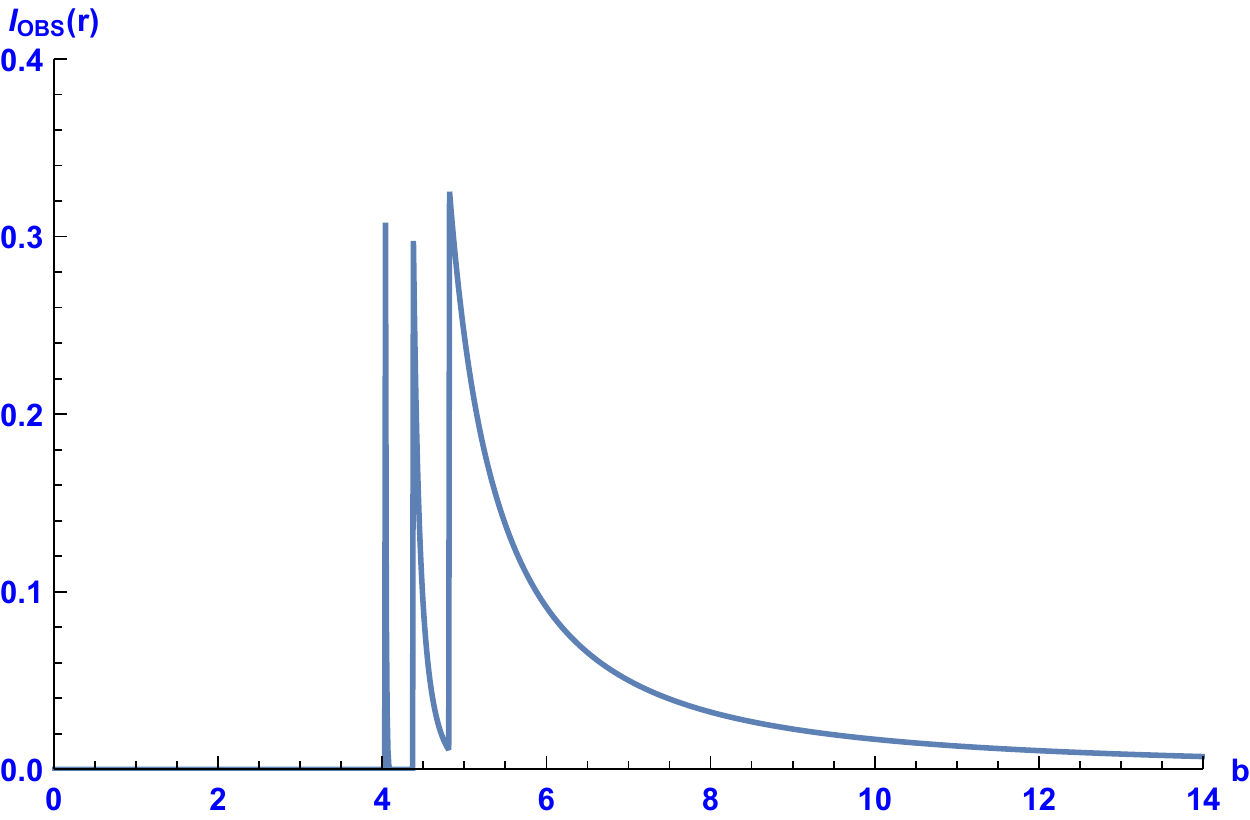}
\includegraphics[width=.28\textwidth]{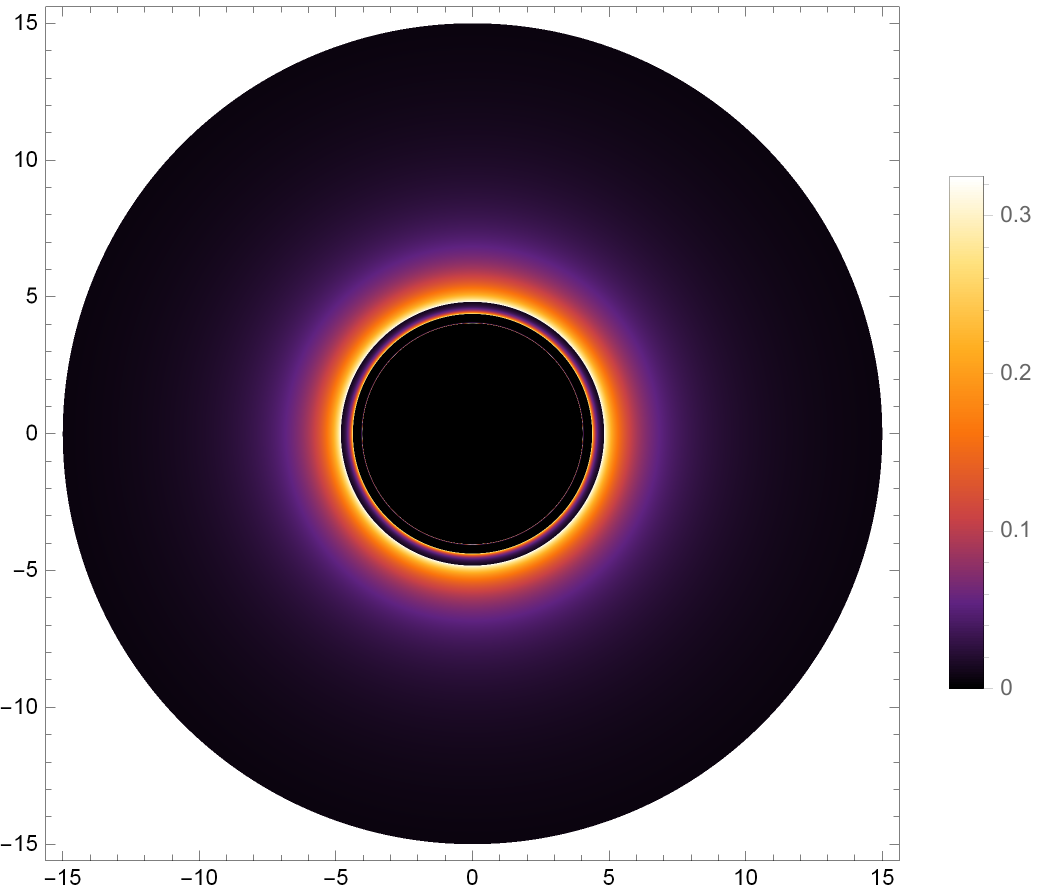}
\includegraphics[width=.35\textwidth]{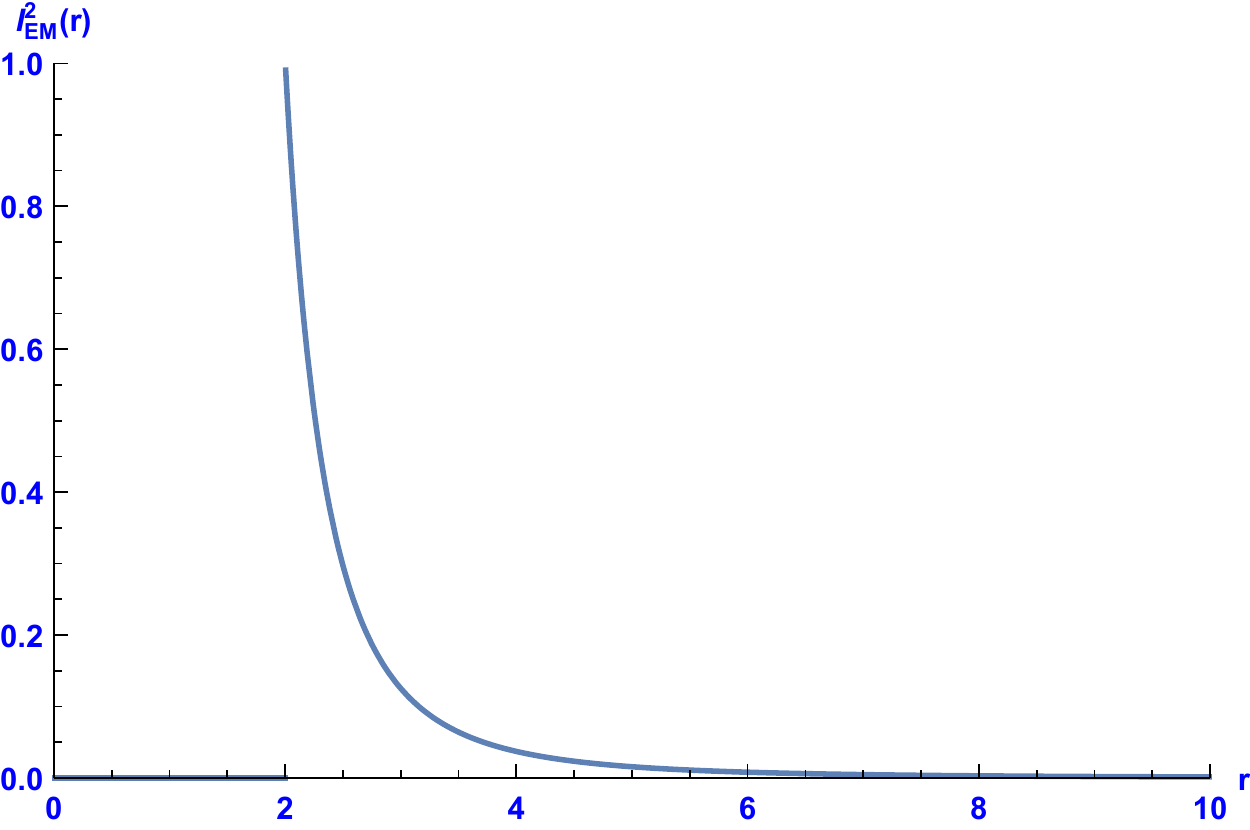}
\includegraphics[width=.35\textwidth]{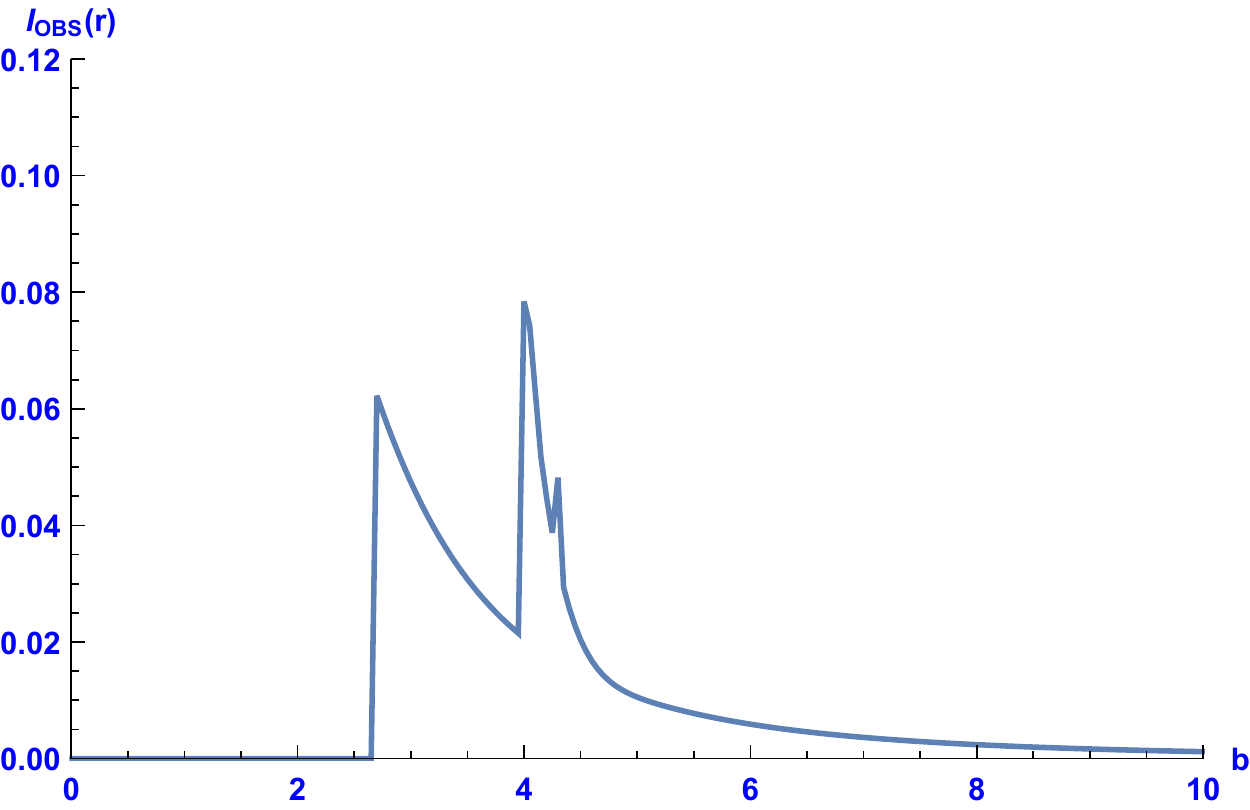}
\includegraphics[width=.28\textwidth]{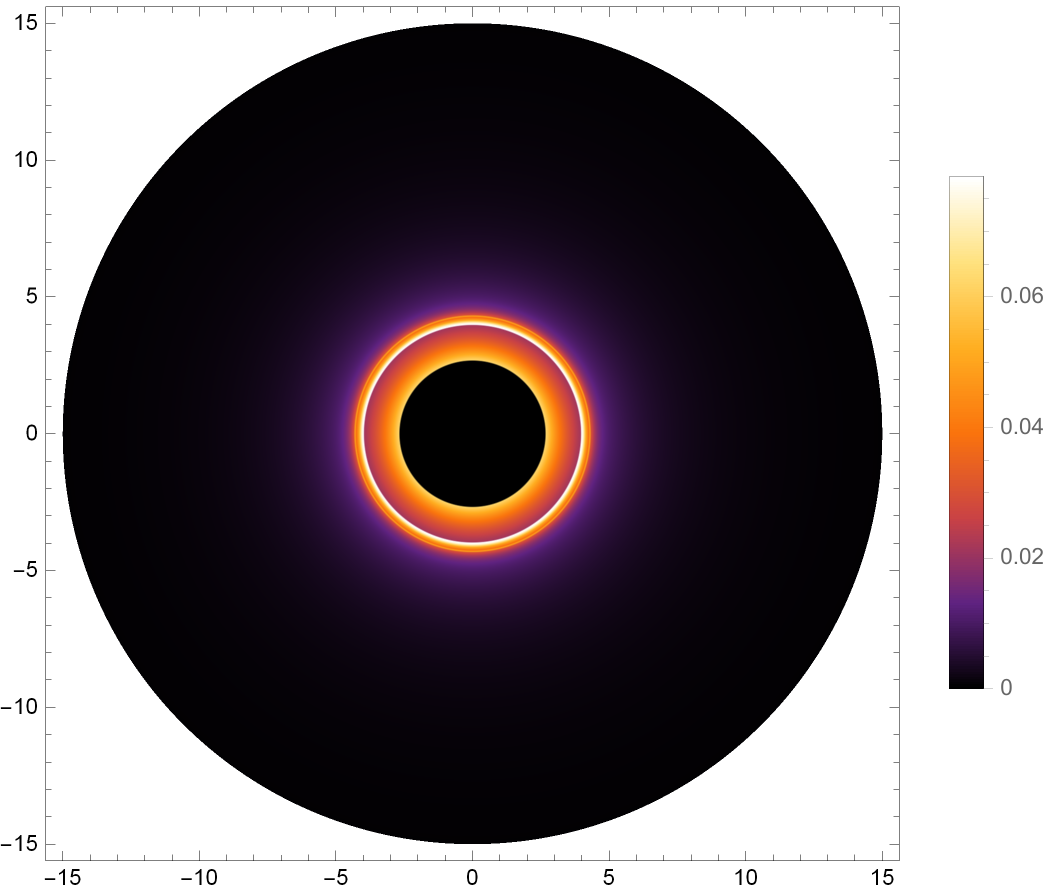}
\includegraphics[width=.35\textwidth]{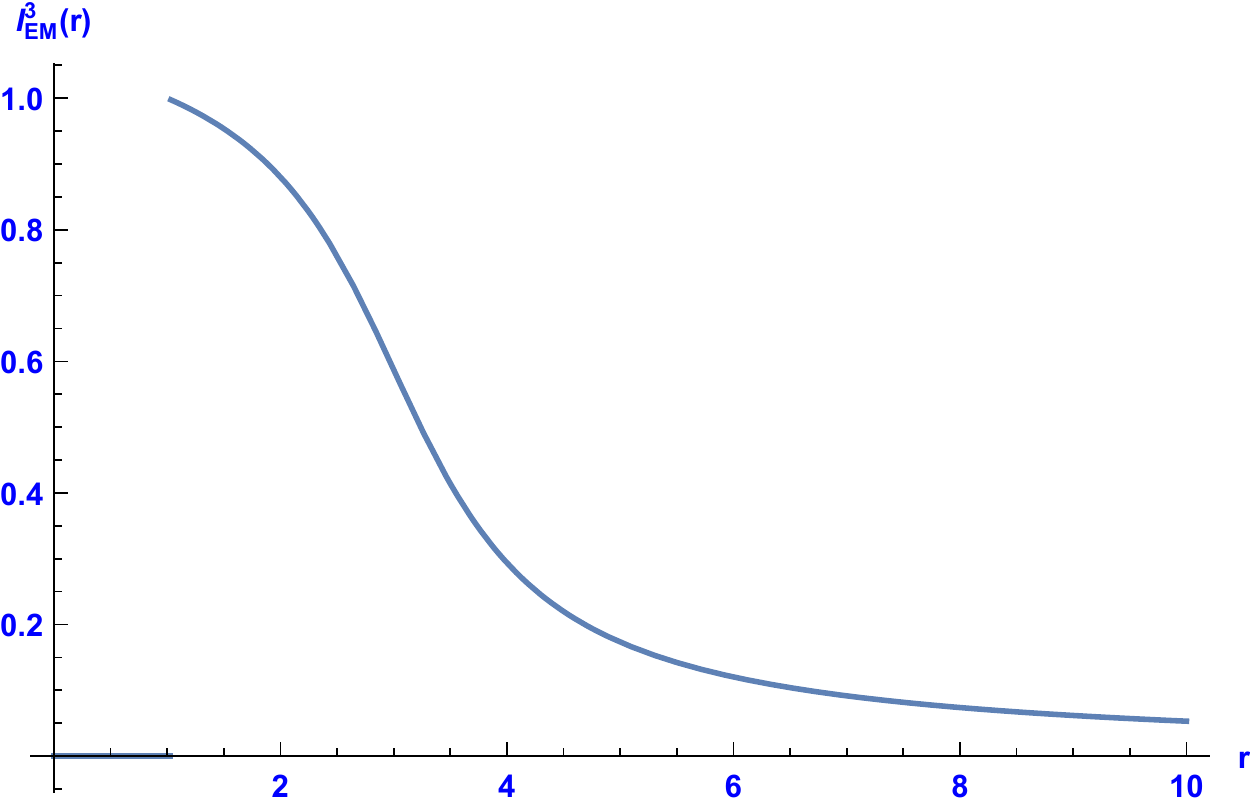}
\includegraphics[width=.35\textwidth]{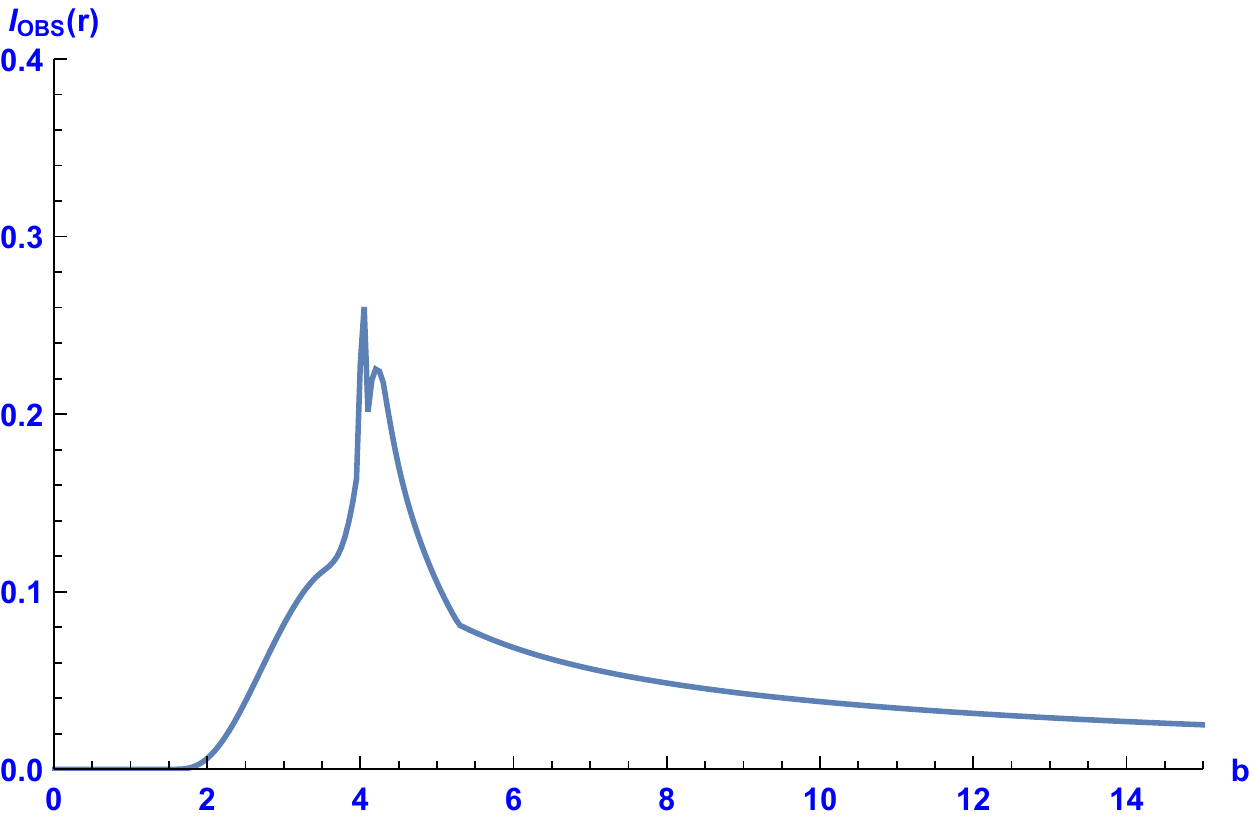}
\includegraphics[width=.28\textwidth]{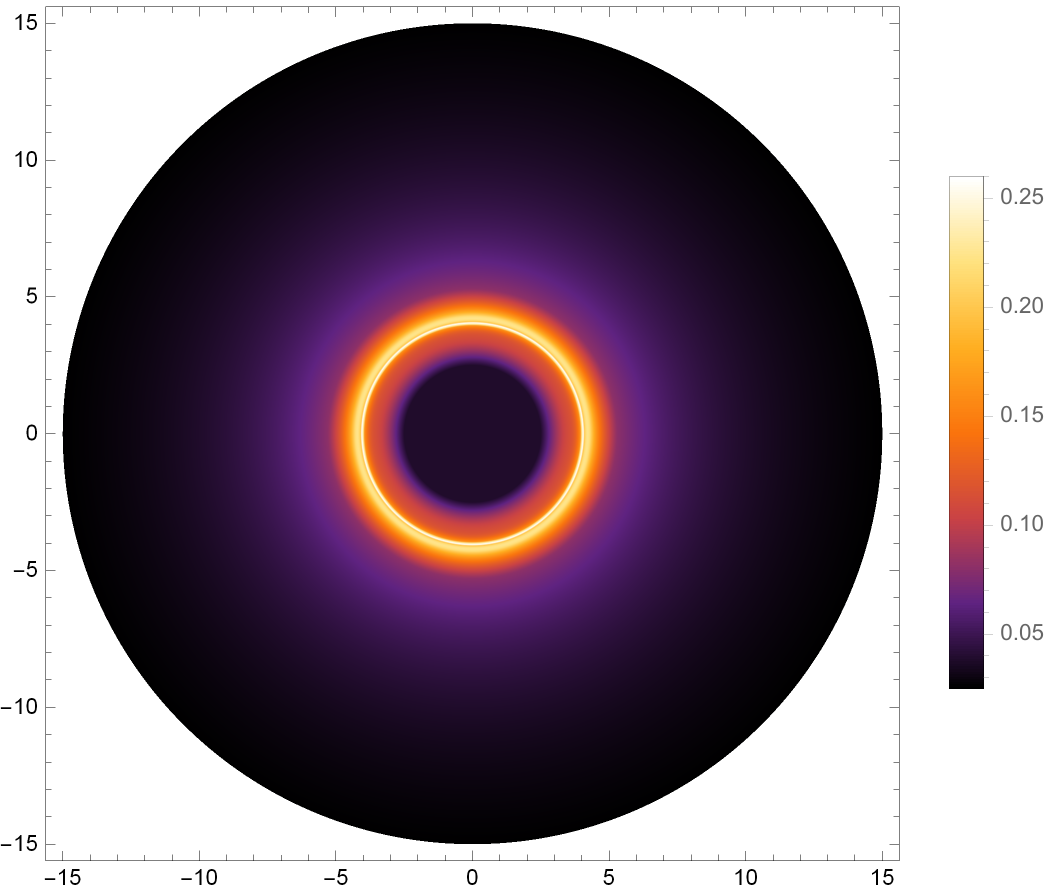}
\caption{\label{fig7}  Observational appearance of the thin disk with different emission profiles  near the black hole ($\theta = 0.049$), viewed from a face-on orientation.}
\end{figure}

From the left column in the first row of Figure 6 and 7, we find that the peak region of emission is near the critical case of the impact parameter $r \sim b_c$ , and then decreases sharply to zero with the increase of radial coordinate $r$. In this case, the photon sphere is in the interior of the emission region. And meanwhile, due to the influence of gravitational lensing effect in the observation process, we clearly see that there are peaks of observed intensity in the region of photon  and lensing ring, and the two peaks are separated independently. However, the peak value of observation intensity of the photon  and lensing ring are not only smaller than those of the direct emission, but also their observation area is very narrow. As a result, the observed intensities is dominated by the direct emission, with the lensing ring emission providing only a small contribution to the total flux and the photon ring providing a negligible contribution. In the column on the right, it shows that the main optical appearance is provided by direct emission, and the photon ring is very difficult to detect.

For the second row, the emission peak  in the left column is near the photon sphere $r\sim r_p$. The results of the middle column show that the direct emission will make the observation intensity reach a peak at first, and then show a gentle attenuation trend with the increase of $r$. More importantly, the region  of photon and the lensing ring almost coincide, which makes the total observation intensity of this area  significantly improved. Hence, the observation intensity will reach a new peak due to the contribution of photon ring, lensing ring and direct emission. However, the photon and lensing ring are highly demagnetized and confined to a narrow region. In other words, the direct emission is still dominant in observations. It can be seen from the right column that the lensing ring has some contribution to the total observation intensity, but the mainly contribution still is the direct emission.

When the emitted region has been extended to the position near the event horizon $r_e$, as shown in the left column of the third row. Moreover, the lensing ring,  photon ring and direct emission  are overlapped in a larger range. The observation intensity increases gently from the position slightly larger than the event horizon, but suddenly increases sharply and reaches a peak in the emission region of the photon ring. Then, due to the contribution of the lensing ring emission, the observation intensity will reach a larger peak. After that, the observation intensity began to decline gently. In this case, the contribution of the lensing ring to the total observation intensity is more obvious than that of the first two models. In addition, the optical appearance shows a bright ring, which is composed of the  contribution by  the lensing ring,  photon ring and direct emission.

Comparing Figure 6 and 7, we can find that the change of noncommutative parameters will have a direct impact on observation in this spacetime. Firstly, with the increase of noncommutative parameters, the radius of photon  and lensing ring decreases, but the area  occupy becomes wider.  It shows that the  optical appearance of model III, the bright ring when $\theta=0.049$ is wider than that when $\theta=0.01$. However, for all three emitted functions, when the noncommutative parameter increases, the maximum observation intensity decreases. In particular, when $\theta = 0$ (the Schwarzschild spacetime \cite{Gralla:2019xty}), the observation intensity is significantly higher than that case of $\theta= 0.049$ (the noncommutative spacetime). Therefore, these results can provide an effective help to distinguish the noncommutative Schwarzschild black hole from the Schwarzschild black hole in spacetime.

\section{Shadows of the noncommutative Schwarzschild black hole with spherical accretion}
In this section, our main goal is to study the image of the noncommutative Schwarzschild black hole with a spherical accretion, which can be regarded to be optically thin. Here, we consider two different models,  that is, the static spherical accretion and infalling spherical accretion. More importantly, we want to explore the influence of noncommutative parameters on the spacetime structure and observational characteristics.

We first study the shadow image and photon sphere of a black hole wrapped by static spherical accretion in the noncommutative spacetime. That is, we assume that the  accretion is static around the black hole. For the observer at infinity, the specific intensity of observation (measured usually in $\rm erg s^{-1} cm^{-2} str^{-1} Hz^{-1}$) is determined by \cite{Jaroszynski:1997bw,Bambi:2013nla}
\begin{align}
{I}({\nu_{obs}})=\int _{\gamma }g^3 j( \nu _e) dl_{{prop}}, \label{Eq4.1}
\end{align}
and
\begin{align}
g=\frac{\nu _{obs}}{\nu _e}, \label{Eq4.2}
\end{align}
among them, $g$, $\nu _e$  and  $\nu _{obs}$ are redshift factor, the photon frequency and  observed photon frequency, respectively. In the rest frame of the emitter,  $j( \nu _e)$ is the emissivity per unit volume, and $dl_{{prop}}$ is the infinitesimal proper length.  In the noncommutative Schwarzschild spacetime, the redshift factor has a form
\begin{align}
g=\frac{\nu _{obs}}{\nu _e}=A(r)^{1/2}, \label{Eq4.3}
\end{align}
Then, we can consider that the radiation of light is not only monochromatic, but also monochromatic $\nu_f$ is the fixed frequency,. When the emission has a $1 / r^2$ radial profile \cite{Bambi:2013nla}, we can get the following equation
\begin{align}
j\left(\nu _e\right)\propto  \frac{\delta \left(\nu _e-\nu _f\right)}{r^2}, \label{Eq4.4}
\end{align}
where $\delta$ stands for the delta function. In this spacetime, we can also obtain
\begin{align}
{dl}_{{prop}}=\sqrt{\frac{1}{A(r)}{dr}^2+r^2{d \varphi }^2}.\label{Eq4.5}
\end{align}
Specially, we can take another form of the above equation, which is
\begin{align}
{dl}_{{prop}}=\sqrt{\frac{1}{A(r)}+r^2\left(\frac{{d\varphi }}{{dr}}\right)^2}{dr}. \label{Eq4.6}
\end{align}
Under the help of the equations (\ref{Eq4.3}), (\ref{Eq4.4}) and (\ref{Eq4.5}), we can get the specific intensity observed by a static observer at infinity is
\begin{align}
I(\nu_{obs})=\int _{\gamma }\frac{A(r)^{3/2}}{r^2}\sqrt{\frac{1}{A(r)}+r^2\left(\frac{{d\varphi }}{{dr}}\right)^2}{dr}. \label{Eq4.7}
\end{align}
Based on the above equation, we can explore the shadow image and corresponding intensity of black hole surrounded by static accretion model in noncommutative spacetime. It is worth mentioning that we want to study the influence of the change of noncommutative parameters on the observation intensity and make a comparison with Schwarzschild spacetime. In Figure 8, we depict the observed specific intensity at spatial infinity when the noncommutative parameter $\theta$ takes different values. At the same time, in Figure 9, we also show these properties of the black hole shadow image and its brightness in  spacetime with the different noncommutative parameters $\theta$.

\begin{figure}[h]
\centering 
\includegraphics[width=0.55\textwidth]{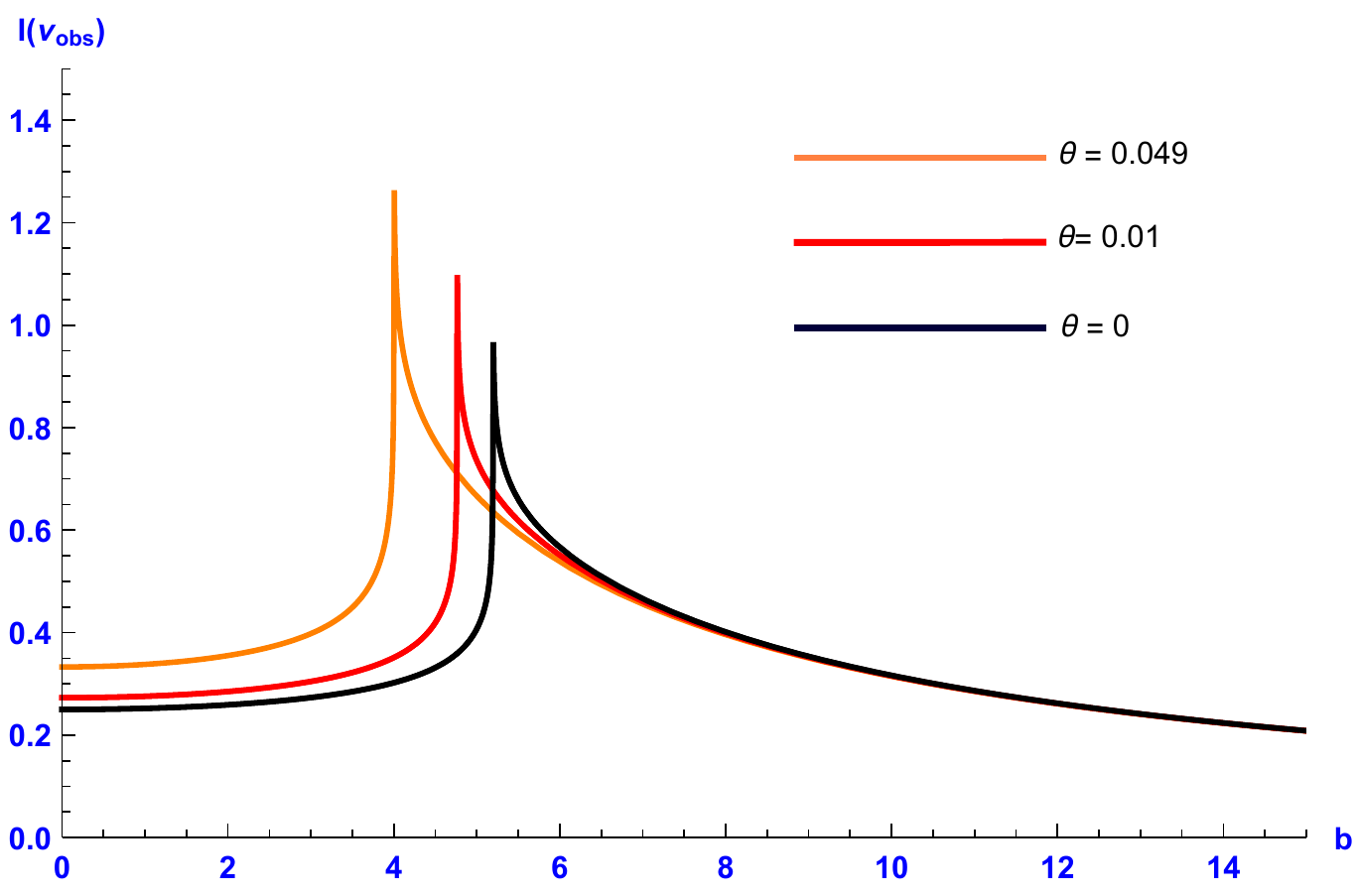}
\caption{\label{fig8} For different noncommutative parameter values $\theta$, the observed specific intensity at spatial infinity with $M=1$.  }
\end{figure}
As illustrated in Figure 1, with the increase of impact parameter $b$, the observed intensity $I(\nu_{obs})$ began to increase gently  until it reached the peak at $b_c$, and then showed a trend of attenuation.

\begin{figure}[h]
\centering 
\includegraphics[width=.325\textwidth]{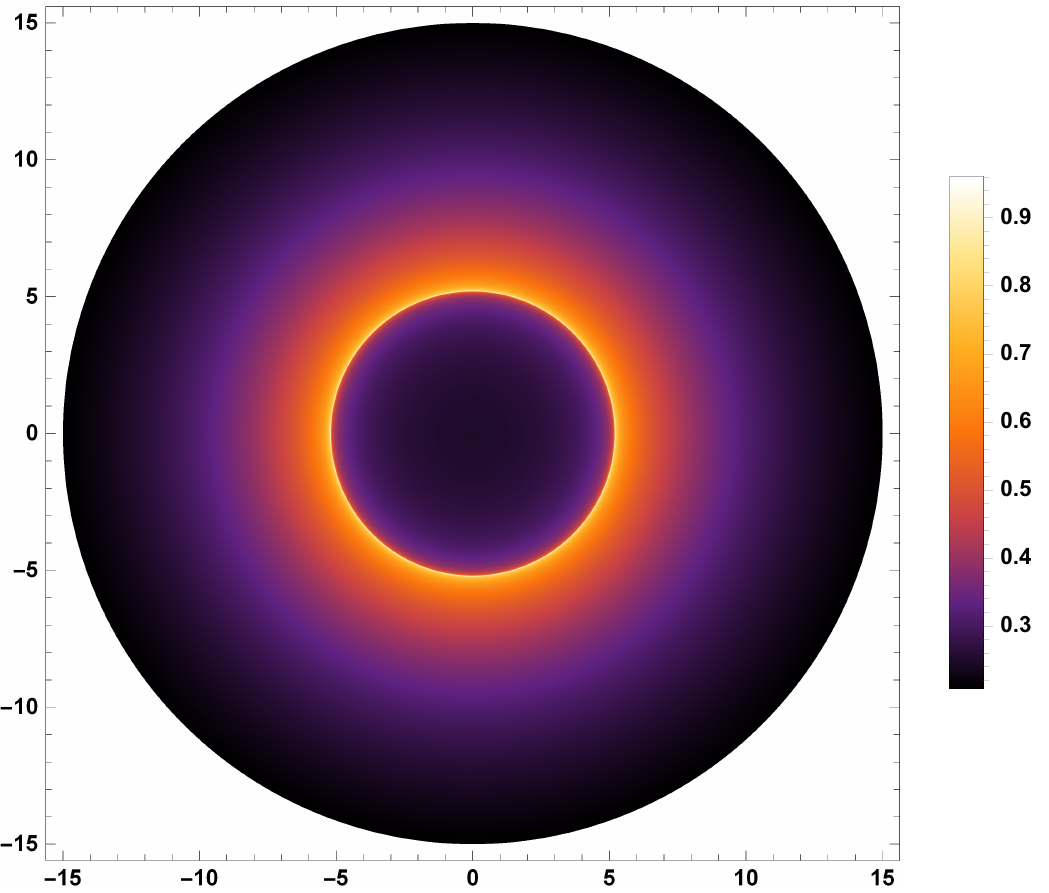}
\hfill
\includegraphics[width=.325\textwidth]{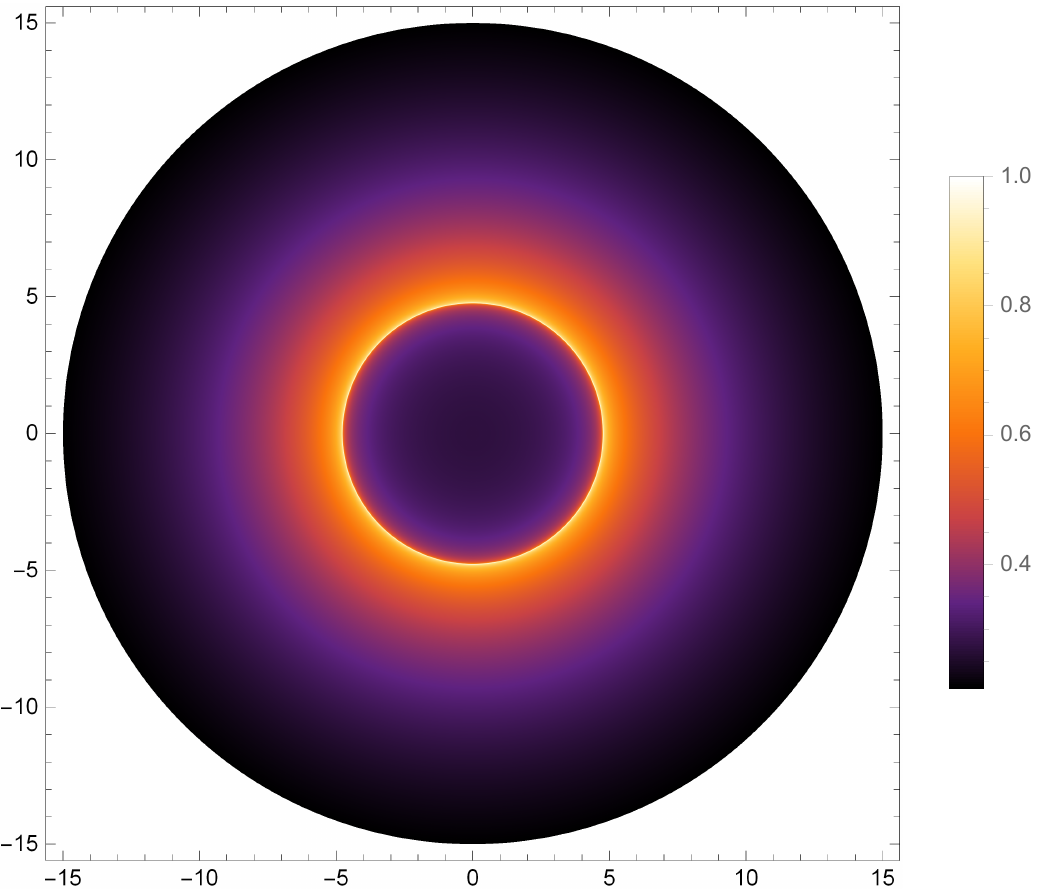}
\hfill
\includegraphics[width=.325\textwidth]{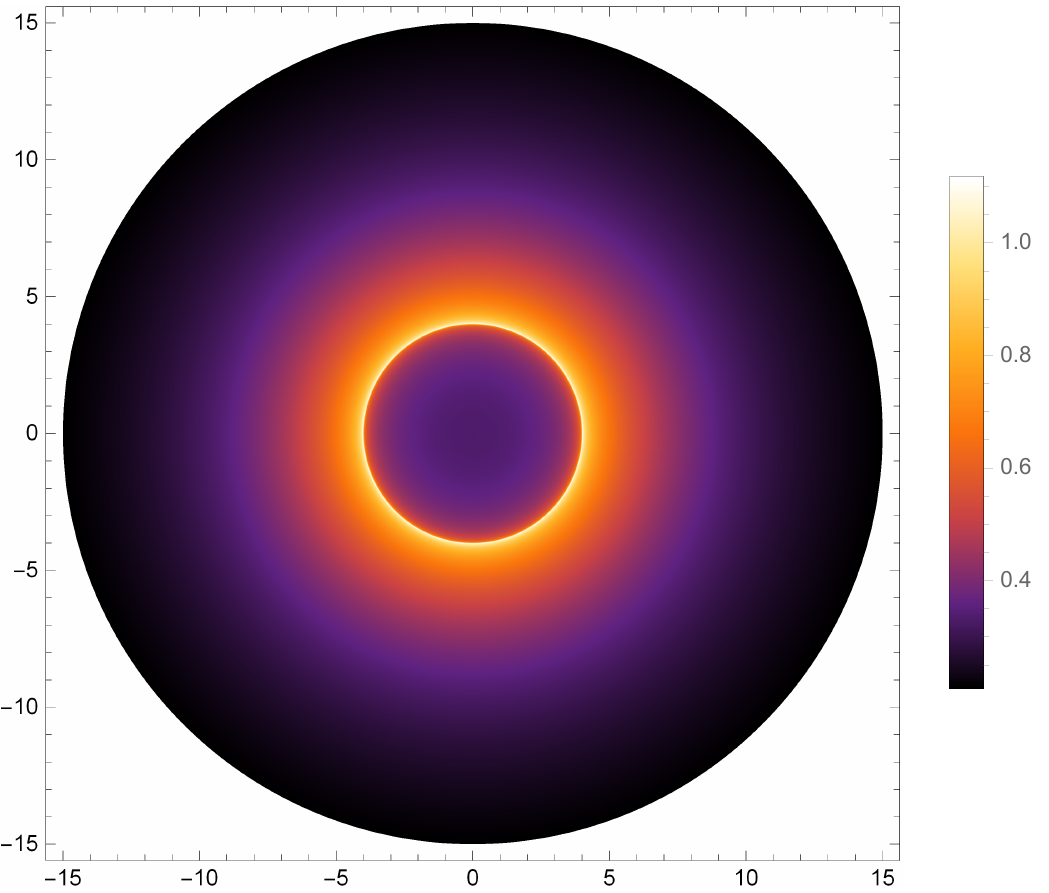}
\caption{\label{fig9}  Images of the black hole shadow with static spherical accretion  for the noncommutative parameters $\theta$, in witch $\theta=0$ (left panel), $\theta=0.01$ (middle panel) and $\theta=0.049$ (right panel).  }
\end{figure}
From figure 8, we can clearly see that the specific intensity $I(\nu_{obs})$  will increase sharply with the increase of the impact parameter $b$, and reach the peak value in the critical case $b\sim b_c$, which is applicable to  different values of the noncommutative parameter $\theta$. When the impact parameter $b$ is larger than the critical condition and continues to increase $b>b_c$, the specific intensity $I(\nu_{obs})$ shows a decreasing trend. And, when $b$ tends to infinity $b\sim \infty$, the observation intensity will be infinitely close to zero $I(\nu_{obs})\sim 0$. Moreover, we find that the change of noncommutative parameter will affect the observation intensity, which is shown in that the increase of noncommutative parameter will increase the peak value of observation intensity. In other words, the larger the noncommutative parameter is, the brighter the optical appearance image will be seen by the observer at infinity. Obviously, these features are also reflected in Figure 9, the  observed optical appearance in noncommutative spacetime ($\theta=0.049$ or $\theta=0.01$) is much brighter in Schwarzschild spacetime ($\theta=0$). However, the radius of the photon sphere of the noncommutative Schwarzschild black hole is obviously smaller than that of the Schwarzschild black hole. In addition, the inner region of the photon sphere in Figure 9 is not completely black, and there is little brightness near the photon sphere, which is caused by a small part of radiation escaping from the black hole.

Now, we will study the spherical accretion around the noncommutative black hole, and the radiative gas moves radially toward the black hole. It is a dynamic model, the equation (\ref{Eq4.1}) is still applicable, but the corresponding redshift factor becomes
\begin{equation}
g=\frac{\mathcal{K}_{\rho } u _0^{\rho }}{\mathcal{K}_{\sigma } u _e^{\sigma }}, \quad  \mathcal{K}^{\mu }=\dot{x}_{\mu },   \label{EQ4.8}
\end{equation}
where $\mathcal{K}^{\mu }$  is the four-velocity of the photon and $u_0^{\mu }=(1,0,0,0)$ is the four-velocity of the static observer. In addition, the quantity $u_e^{\mu }$ is the four-velocity of the  infalling accretion, which has a form
\begin{equation}
u_e^t={A (r)}^{-1},\quad u_e^r=-\sqrt{{1-A (r)}}, \quad u_e^{\theta }=u_e^{\varphi }=0.   \label{EQ4.9}
\end{equation}
From the null geodesic, we can obtain the four-velocity of photon, that is
\begin{equation}
\mathcal{K}_t=\frac{1}{b}, \quad \frac{\mathcal{K}_r}{\mathcal{K}_t}= \pm \frac{1}{A (r)}\sqrt{1-A(r) \frac{b^2}{r^2}},  \label{EQ4.10}
\end{equation}
Considering that photon may approach or escape from the black hole, the above equation is preceded by a sign $\pm$. Then, for the case of infalling spherical accretion, the redshift factor can be written as follows
\begin{equation}
{g}_i=[u_e^t+(\frac{\mathcal{K}_r }{\mathcal{K}_e})u_e^r] ^{-1}. \label{EQ4.11}
\end{equation}
And, the proper distance can be defined by
\begin{equation}
d l_{p}=\mathcal{K}_\mu u_e ^\mu d\lambda=\frac{\mathcal{K}_t}{{g}_i |\mathcal{K}_r|} d r. \label{EQ4.12}
\end{equation}
Similarly, we also assume that the specific emissivity is monochromatic. Therefore, the observation intensity in the case of infalling spherical accretion can be expressed as
\begin{equation}
{I}^\ast({\nu_{obs}})\propto \int _{\gamma }\frac{{g}_i^3  \mathcal{K}_t dr}{r^2 |\mathcal{K}_r|}.\label{EQ4.13}
\end{equation}
Using the  equation (\ref{EQ4.13}), we can study the influence of noncommutative parameters on black hole shadow image and its brightness distribution, as shown in Figure 10 and Figure 11.
\begin{figure}[h]
\centering 
\includegraphics[width=0.65\textwidth]{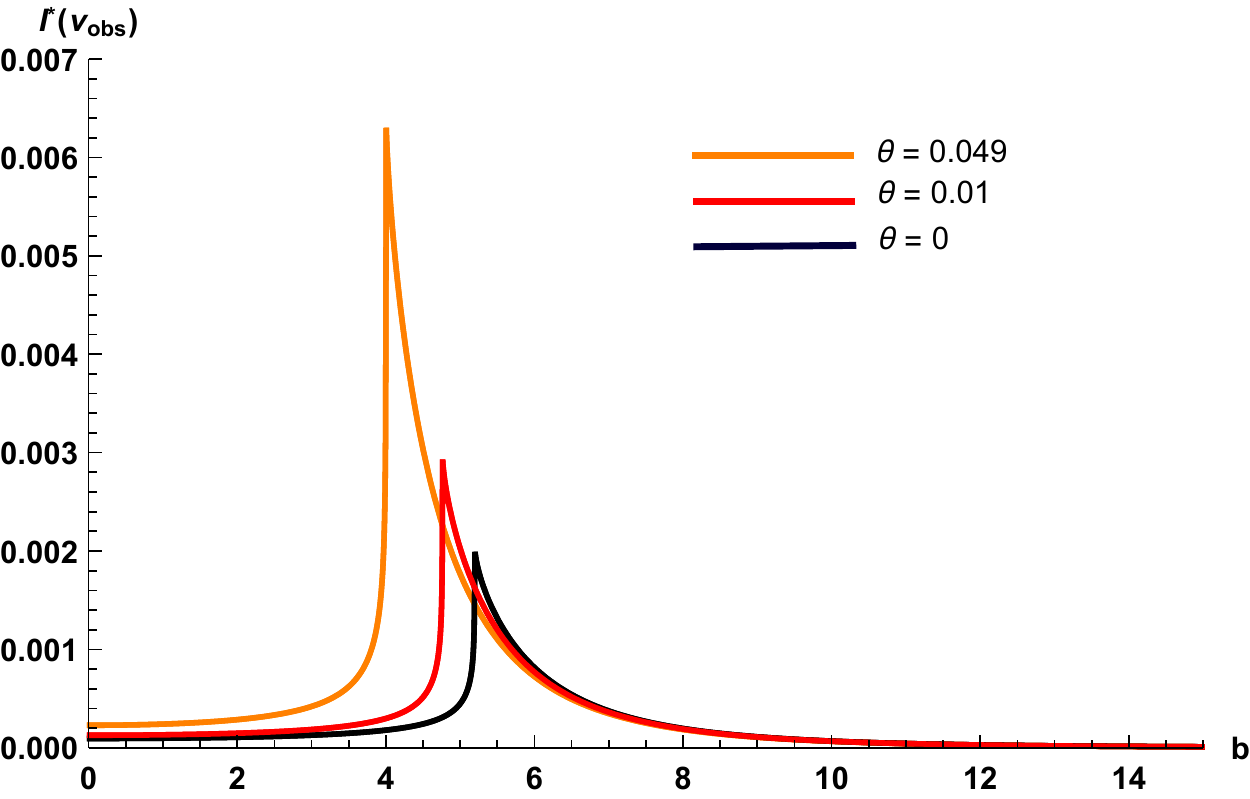}
\caption{\label{fig10} The observed specific intensity at spatial infinity for different noncommutative parameter values $\theta$ with $M=1$.  }
\end{figure}

\begin{figure}[h]
\centering 
\includegraphics[width=.325\textwidth]{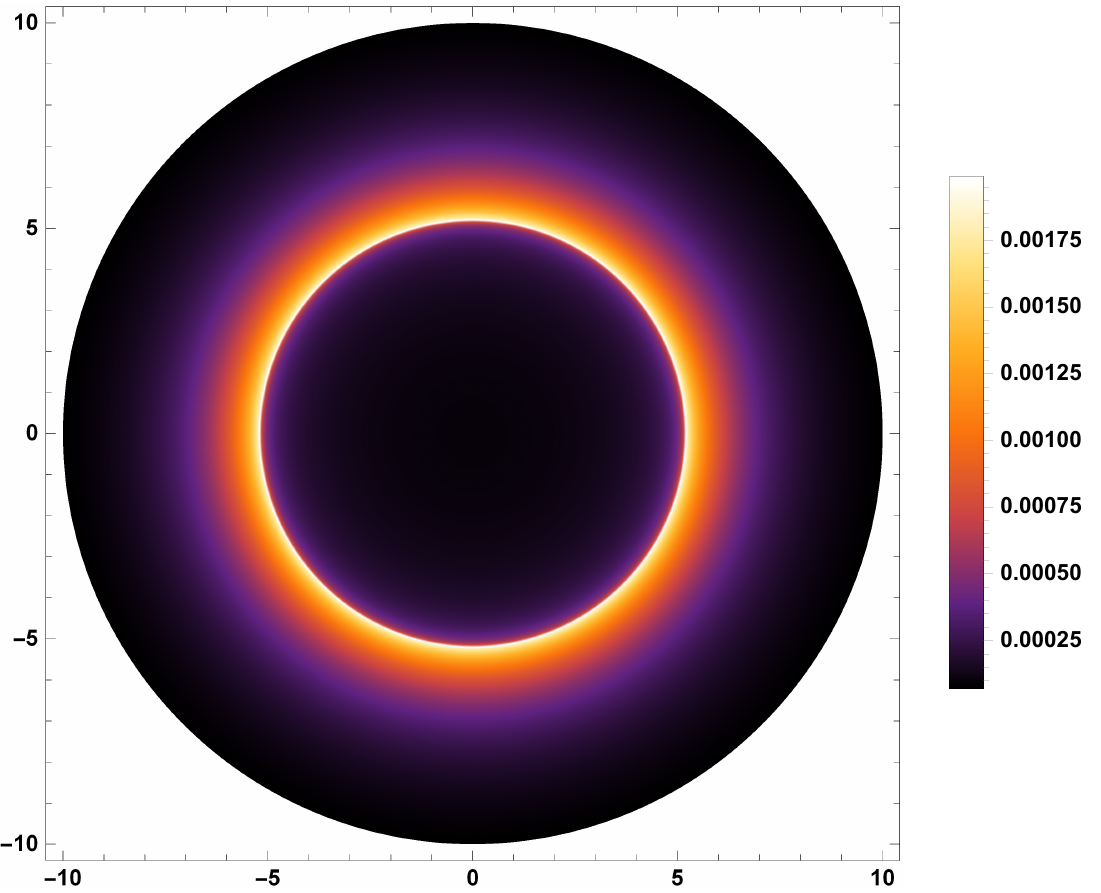}
\hfill
\includegraphics[width=.325\textwidth]{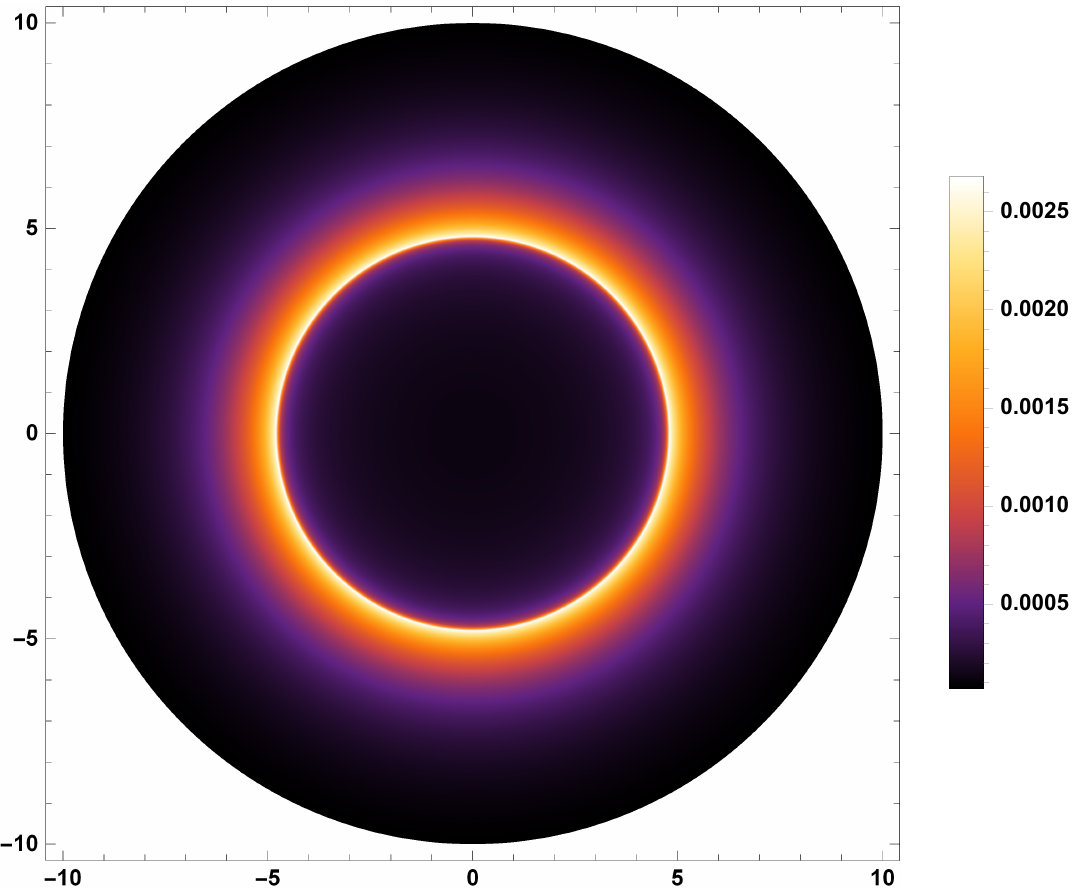}
\hfill
\includegraphics[width=.325\textwidth]{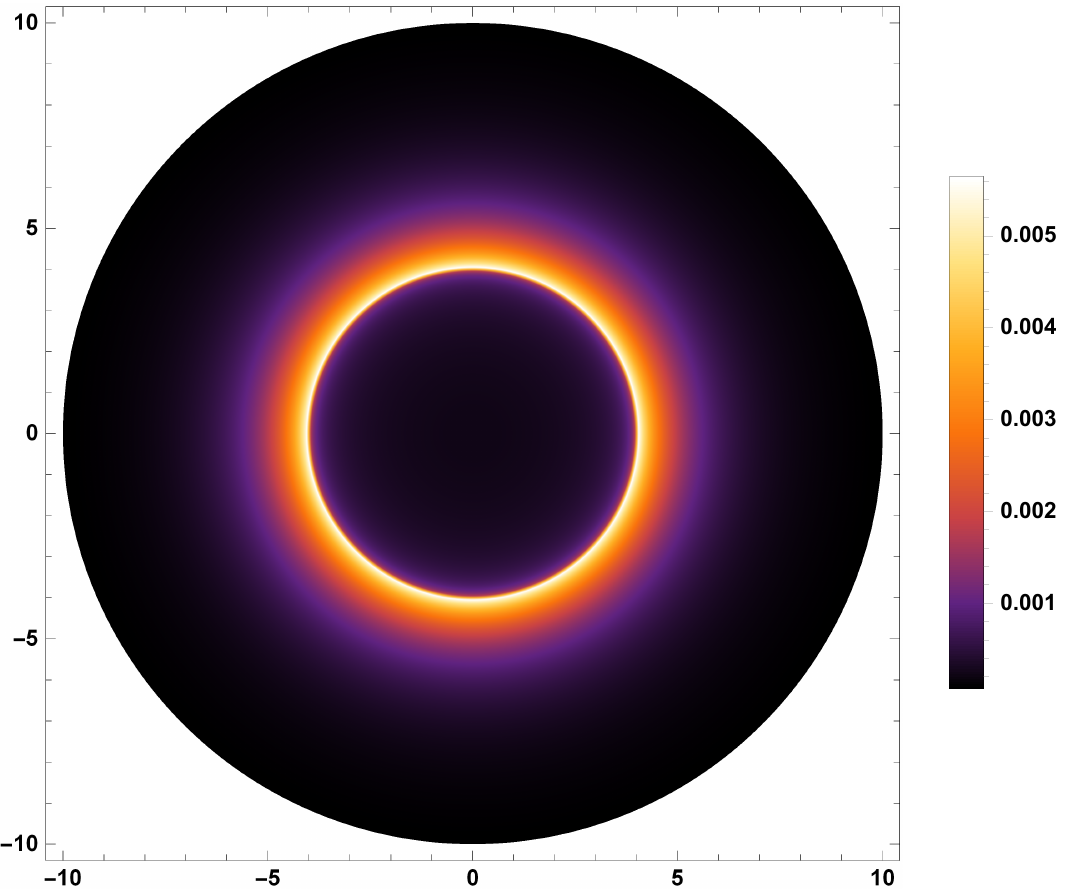}
\caption{\label{fig11}  Images of the black hole shadow with infalling spherical accretion  for the noncommutative parameters $\theta$, in witch $\theta=0$ (left panel), $\theta=0.01$ (middle panel) and $\theta=0.049$ (right panel).  }
\end{figure}
For different state parameters $\theta$, the maximum observed intensity is still at the critical position of the impact parameter $(b\sim b_c)$. When $b < b_c$, the observation intensity  increases with the increase of $b$ until it reaches the peak. However, when $b > b_c$, the observation intensity tends to decrease with the increase of $b$. In addition, the radius of black hole shadow image decreases with the increase of noncommutative parameter, but the observed brightness increases. Both Figure 10 and Figure 11 show that the observed luminosity at $\theta = 0.049$ is significantly higher than that at $\theta = 0.01$ and $\theta = 0$ (the Schwarzschild spacetime). It is worth noting that the central region of the infalling spherical accretion is much darker than that of the static accretion case, which is due to the Doppler effect.

\section{Conclusions and discussions}
\label{conclusion}
In this paper, we mainly study the shadow and observation characteristics of noncommutative Schwarzschild black holes wrapped by three accretion models, and  then explore the influence of noncommutative parameters on the observation appearance and spacetime geometry of black holes.
Based on the  null geodesic of the noncommutative black hole, we study the effective potential and photon orbits in this spacetime. Due to the change of noncommutative parameters, the spacetime structure changes, which also leads to the change of corresponding physical parameters. Specifically, the event horizon $r_e$, the radius $r_p$ and impact parameter $b_c$ of photon sphere are all decreased with the increase of the noncommutative parameter $\theta$, but the effective potential increased. In order to study the observational appearance and intensity around a noncommutative black hole surrounded by accretion matter, we mainly take three optical and geometric accretion models as examples to present the observational appearance of the noncommutative black hole.

When the noncommutative Schwarzschild black hole is surrounded by an optically and geometrically thin accretion disk in the frame of the static worldlines, we study the ray trajectories near the black hole according to the total number of orbits $n=\frac{\varphi}{2 \pi}$. Those ray trajectories can be divided into the direct emission, lensing ring and photon ring. In this sense, it turns out that there is not only a dark central shadow area, but also the photon rings and lensing ring in the outer area of the shadow. Intriguingly, by studying the first three transfer functions, we find that the lensing ring is highly demagnetized, while the photon ring is extremely demagnetized. In addition, we employ three typical toy model functions of the emission profile to further study the observational appearance of the noncommutative black hole, and then compare the observed specific intensities of direct emission, lensing ring and photon ring. It shows  that  the photon ring is a highly curved light, and it intersects with the disk plane at least three times, but its contribution to the total brightness can be ignored because its area is so narrow. Then, the area occupied by the lensing ring is wider than photon ring, and its demagnetization is not too high in a certain range of the impact parameter, thus making a bigger contribution to the flux of the observed image. But, this contribution is also very small by comparing with the direct emission. Hence, our results show that the direct emissions always dominate the total observed intensity for all three toy-model functions. When the noncommutative parameters increased, the range of lensing ring and photon ring will increase. Also, the observation intensity influenced naturally, i.e., one can see in Figure 7 that there appears a wide and bright annulus in the optical observation appearance of model III. Moreover, in order to get some insight into the effects of these  rings on a realistic observation, we blur the images of the accretion disk, which shown in Figure 12. Here, we mimic the EHT resolution by using a Gaussian filter with standard derivation equal to $1/12$ the field of view\cite{Gralla:2019xty}. Obviously, the characteristics of the lensing ring in the simulation observation results are washed out, and the photon ring can not be observed directly. Indeed, the blurred rings and direct emission are indistinguishable for all three emitted functions, even if they are distinguishable before blurring. Although the characteristics of photon ring and lensing ring are weakened, the third emission mode has a brighter and wider blurred ring compared with the first two emission modes. In this case, the shadow radius is quite different, which means that the location of the direct emission determines the size of the shadow.
\begin{figure}[h]
\centering
\subfigure[model I]{\includegraphics[scale=0.45]{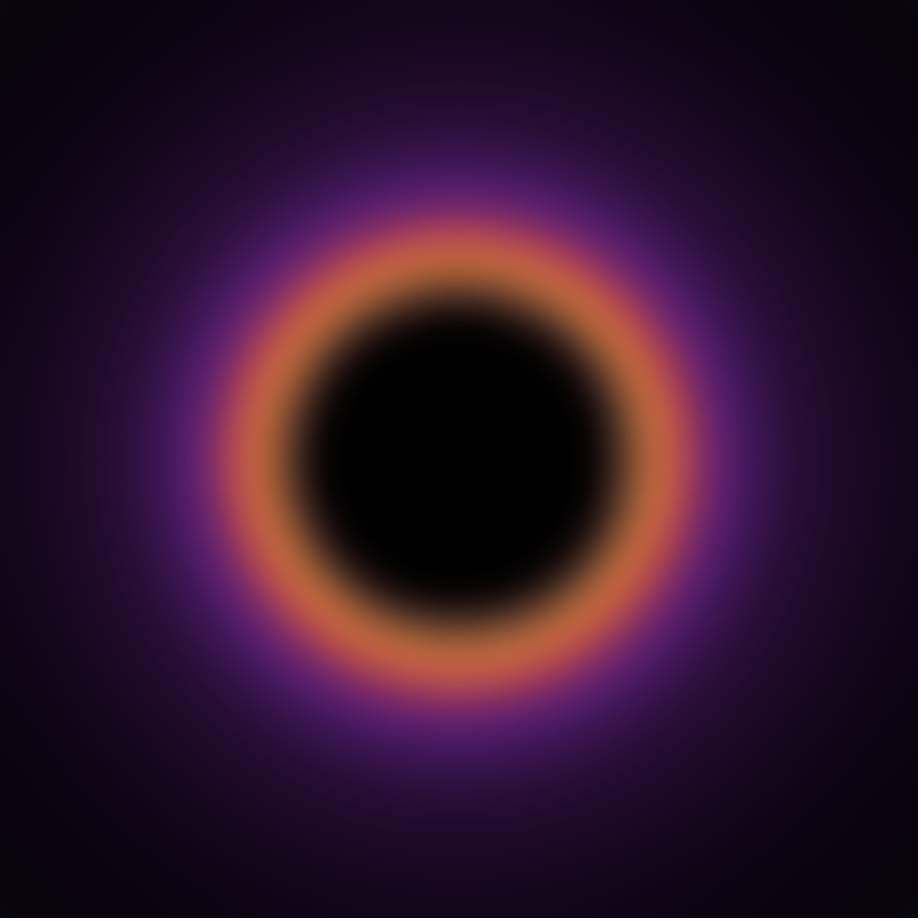}} \qquad
\subfigure[model II]{\includegraphics[scale=0.45]{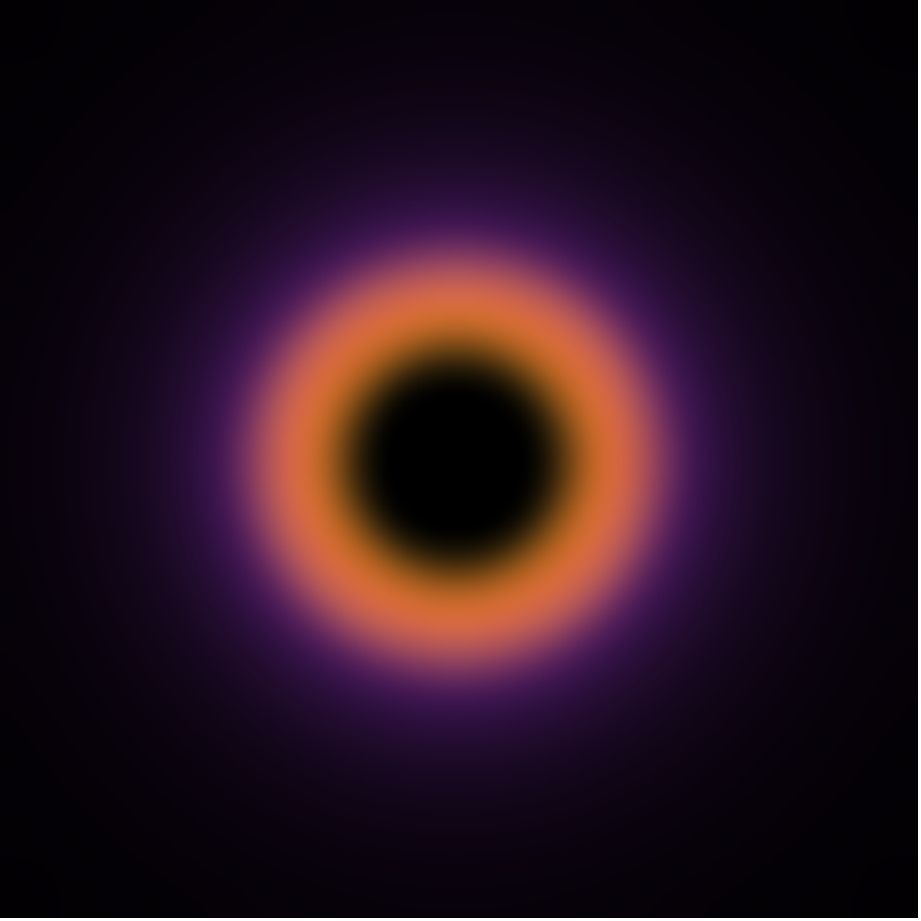}} \qquad
\subfigure[model III]{\includegraphics[scale=0.45]{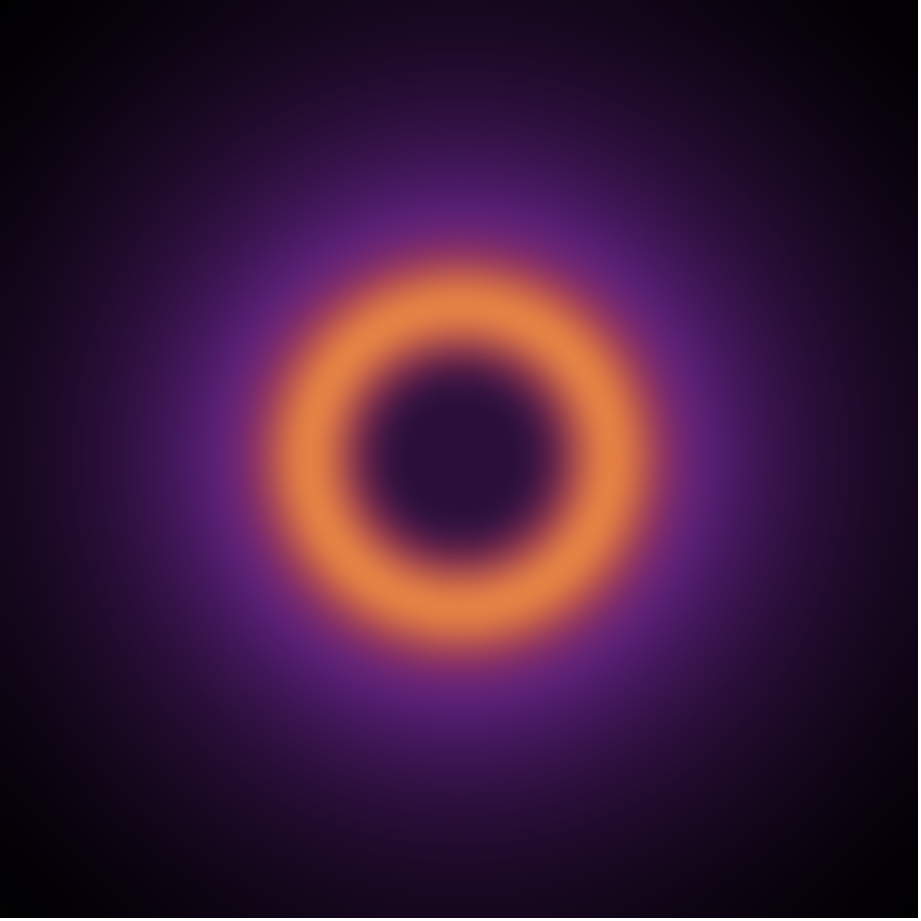}}
\caption{ The blurred images from Figure 7, and The high resolution images are blurred to correspond roughly to the EHT resolution.}
\end{figure}

In our universe, there may still be spherically symmetric accretion around black holes. It is of great significance to further discuss the observational appearance of black holes surrounded by spherical accretion.
For convenience, we here consider only two simple relativistic spherical accretion models, i.e., the static spherical accretion and infalling spherical accretion. In both spherical accretion models, the maximum observed brightness always appears at the impact parameter $b\sim b_c$, and the brightness of the inner area of the shadow is very low, but the observation brightness outside the shadow increases obviously. Moreover, with the increase of the noncommutative parameter $\theta$, the observation intensity of the optical appearance in both two models also increase. However, it is obvious from Figures 8 and 10 that the infalling accretion model increase significantly higher than the static accretion. In addition, due to the Doppler effect, the inner area of the shadow of the infalling accretion is darker than that of the static. This model is considered to be more realistic than the static spherical accretion model since most of the accretion in the universe is dynamic.

\vspace{10pt}

\noindent {\bf Acknowledgments}

\noindent
The authors would like to thank the anonymous reviewer for their helpful comments and suggestions, which helped to improve the quality of this paper. This work is supported  by the National Natural Science Foundation of China (Grant Nos. 11675140, 11705005, 11903025 and 11875095).



\end{document}